\documentclass[12pt,letterpaper]{article}
\usepackage{amsfonts}
\usepackage{amssymb}
\usepackage{amsmath}
\usepackage{amsthm}
\usepackage[round,authoryear]{natbib}
\usepackage[margin=1in]{geometry}
\usepackage{booktabs}
\usepackage{dcolumn}
\newcolumntype{d}[1]{D{.}{.}{#1}} 
\usepackage{graphicx,epstopdf}
\usepackage[T1]{fontenc}
\usepackage{lmodern}
\usepackage{enumitem}
\setlist[enumerate,1]{itemsep=1pt, topsep=1pt, partopsep=0pt, parsep=1pt}
\setlist[enumerate,2]{nosep}
\setlist[itemize,1]{itemsep=1pt, topsep=1pt, partopsep=0pt, parsep=1pt}
\setlist[itemize,2]{nosep}
\usepackage{subfig}
\usepackage{bm}
\usepackage[colorlinks,linkcolor=red,citecolor=blue,pdftex,breaklinks]{hyperref}
\hypersetup{bookmarksopen=true}
\usepackage[nameinlink]{cleveref}
\usepackage{setspace}

\usepackage{color}
\usepackage{longtable}
\usepackage{array}
\newcolumntype{Y}{>{\raggedright\arraybackslash}p{0.8\textwidth}}

\theoremstyle{plain}

\theoremstyle{definition}

\newtheorem{examplex}{Example}
\newenvironment{example}{\pushQED{\qed}\examplex}{\popQED\endexamplex}

\Crefname{assumptionx}{Assumption}{Assumptions} 
\Crefname{examplex}{Example}{Examples} 
\Crefname{remarkx}{Remark}{Remarks} 

\makeatletter
\renewcommand*{\eqref}[1]{\hyperref[{#1}]{\textup{\tagform@{\ref*{#1}}}}}
\makeatother

\makeatletter
\DeclareRobustCommand\citepos
  {\begingroup\def\NAT@nmfmt##1{{\NAT@up##1's}}%
   \NAT@swafalse\let\NAT@ctype\z@\NAT@partrue
   \@ifstar{\NAT@fulltrue\NAT@citetp}{\NAT@fullfalse\NAT@citetp}}
\makeatother

\expandafter
\def \expandafter \normalsize \expandafter{\normalsize \setlength \abovedisplayskip{10pt plus 2pt minus 7pt}}
\expandafter
\def \expandafter \normalsize \expandafter{\normalsize \setlength \abovedisplayshortskip{0pt plus 2pt}}
\expandafter
\def \expandafter \normalsize \expandafter{\normalsize \setlength \belowdisplayskip{10pt plus 2pt minus 7pt}}
\expandafter
\def \expandafter \normalsize \expandafter{\normalsize \setlength \belowdisplayshortskip{5pt plus 2pt minus 3pt}}

\def\bbeta{{\bm\beta}}
\def\biota{{\bm\iota}}
\def\biX{{\bm{X}}}

\def\biM{{\bm{M}}}
\def\biP{{\bm{P}}}
\def\biA{{\bm{A}}}

\def\biH{{\bm{H}}}

\def\bie{{\bm{e}}}
\def\biw{{\bm{w}}}

\def\bis{{\bm{s}}}
\def\biy{{\bm{y}}}
\def\bix{{\bm{x}}}

\def\biu{{\bm{u}}}

\def\bOmega{{\bm{\Omega}}}
\def\bSigma{{\bm{\Sigma}}}

\def\E{{\rm E}}
\def\U{{\rm U}}

\def\tk{\kern 0.08333em}
\def\tn{\kern -0.08333em}
\def\tkk{\kern 0.04167em}
\def\bzero{{\bm{0}}}
\def\bfI{{\bf I}}

\def\th#1{$#1^{\tk{\rm th}}$}

\def\addh#1{\vrule height#1pt depth0pt width0pt}

\DeclareMathOperator{\Tr}{Tr}

\DeclareMathOperator{\var}{Var}

\def\hangpara{\hangindent=\parindent\hangafter=1\noindent}

\newcommand{\FO}[1]{}

\begin{document}
	


\title{Leverage, Influence, and the Jackknife\\ in Clustered
Regression Models:\\ Reliable Inference Using summclust\thanks{
We are grateful to the editor, an anonymous referee, Alexander
Fischer, Rapha\"el Lange\-vin, and seminar participants at York
University and the 2022 and 2023 CEA Annual Meetings for comments. We
are especially grateful to David Drukker for a very insightful
suggestion. MacKinnon and Webb thank the Social Sciences and
Humanities Research Council of Canada for financial support (SSHRC
grants 435-2016-0871 and 435-2021-0396). Nielsen thanks the Danish
National Research Foundation for financial support (DNRF Chair grant
number DNRF154).}}

\author{James G. MacKinnon\thanks{Corresponding author. Address: 
Department of Economics, 94 University Avenue, Queen's University, 
Kingston, Ontario K7L 3N6, Canada. Email:\
\texttt{mackinno@queensu.ca}. Tel.\ 613-533-2293. Fax
613-533-6668.}\\Queen's University\\ \texttt{mackinno@queensu.ca} \and
Morten \O rregaard Nielsen\\Aarhus University\\
\texttt{mon@econ.au.dk} \and Matthew D. Webb\\Carleton
University\\ \texttt{matt.webb@carleton.ca}}

\maketitle
\vskip -12pt

\begin{abstract}
We introduce a new \texttt{Stata} package called \texttt{summclust}
that summarizes the cluster structure of the dataset for linear
regression models with clustered disturbances. The key unit of
observation for such a model is the cluster. We therefore propose
cluster-level measures of leverage, partial leverage, and influence
and show how to compute them quickly in most cases. The measures of
leverage and partial leverage can be used as diagnostic tools to
identify datasets and regression designs in which cluster-robust
inference is likely to be challenging. The measures of influence can
provide valuable information about how the results depend on the data
in the various clusters. We also show how to calculate two jackknife
variance matrix estimators efficiently as a byproduct of our other
computations. These estimators, which are already available in
\texttt{Stata}, are generally more conservative than conventional
variance matrix estimators. The \texttt{summclust} package computes
all the quantities that we discuss.

\vskip 2pt

\medskip \noindent \textbf{Keywords:} summclust, clustered data, 
cluster-robust variance estimator, CRVE, grouped data, high-leverage 
clusters, influential clusters, jackknife, partial leverage, robust 
inference.

\medskip \noindent \textbf{JEL Codes:} C10, C12, C21, C23, C87.

\end{abstract}

\clearpage
\onehalfspacing

\section{Introduction}
\label{sec:intro}


It is now standard in many fields of economics and other disciplines
to employ cluster-robust inference for the parameters of linear
regression models. In the most common case, each of the $N$
observations is assigned to one of $G$ disjoint clusters, which might
correspond to, for example, families, schools, villages, hospitals,
firms, industries, years, cities, counties, or states. The assignment
of observations to clusters is assumed to be known, and observations
in different clusters are assumed to be independent, but any pattern
of heteroskedasticity and/or dependence is allowed within each
cluster. Under these assumptions, a cluster-robust variance matrix, or
CRVE, yields asymptotically valid $t$-tests, Wald tests, and
confidence intervals. However, even when $N$ is very large, the
resulting inferences may be unreliable when $G$ is not large or the
clusters are not sufficiently homogeneous.


The literature on cluster-robust inference has grown rapidly in recent
years. \citet*{CM_2015} is a classic survey article. \citet*{CGH_2018}
surveys a broader class of methods for dependent data.
\citet*{MNW-guide} is a comprehensive guide to empirical practice. As
it discusses, there are two situations in which cluster-robust
$t$-tests and Wald tests are at risk of over-rejecting to an extreme
extent, even when $G$ is not small. The first is when one or a few
clusters are much larger than the rest, and the second is when the
only ``treated'' observations belong to just a few clusters;
\citet*{DMN_2019} discusses the first case, and
\citet*{MW-JAE,MW-TPM,MW-EJ} discuss the second. In both of these
cases, one cluster (or a few of them) has high leverage, in the sense
that omitting this cluster has the potential to change the OLS
estimates substantially. When that actually happens, a cluster is said
to be influential.


The concepts of leverage and influence are normally applied at the
observation level \citep*{BKW_1980}, but they are equally applicable
at the cluster level. Just as high-leverage observations can make
heteroskedasticity-robust inference unreliable \citep{Chesher_1989},
so too can high-leverage clusters make cluster-robust inference
unreliable. Just as highly influential observations may lead us to
suspect that there is something wrong with the model or the data, so
too may highly influential clusters. Any situation in which a few
clusters have high leverage or high influence should be worrying.

There are at least two different concepts of leverage. The usual one
focuses on fitted values or, equivalently, residuals. A cluster is
said to have high leverage if removing it has the potential to change
the fitted values for that cluster by a lot. The second concept is
partial leverage \citep*{CW_1980}. A cluster is said to have high
partial leverage for the \th{j} coefficient if removing that cluster
has the potential to change the estimate of the \th{j} coefficient by
a lot. We discuss both concepts in \Cref{sec:influential}. 


Whether a cluster has high leverage, high partial leverage, or is
influential can depend on the sample in rather complicated ways. We
provide a new \texttt{Stata} package called \texttt{summclust} that
implements computationally-efficient ways to identify high-leverage
and influential clusters and provides a number of statistics that
collectively summarize the cluster structure of the dataset. These can
be useful for detecting cases in which cluster-robust inference may
not be reliable. Our leverage and influence calculations also allow us
to compute two cluster jackknife variance matrix estimators, which we
refer to as CV$_{\tn3}$ and CV$_{\tn3{\rm J}}$, at little additional
cost. These estimators are already available in \texttt{Stata} by
using either the \texttt{vce(jackknife)} option or the
\texttt{jackknife} prefix. Recent work
\citep*{Hansen-jack,MNW-bootknife} suggests that CV$_{\tn3}$ and
CV$_{\tn3{\rm J}}$ generally perform better in finite samples than
more widely-used CRVEs; see also \Cref{sec:sims}.


The remainder of the paper is organized as follows. The next section
begins with a brief review of cluster-robust inference for linear
regression models. Then \Cref{sec:influential} introduces our new
measures of leverage, partial leverage, and influence at the cluster
level. \Cref{sec:jack} shows how our results can be used to compute
the CV$_{\tn3}$ and CV$_{\tn3{\rm J}}$ jackknife variance matrix
estimators. \Cref{sec:report} discusses what quantities are reported
by \texttt{summclust} and should, at least in some cases, be reported
by the investigator.


\Cref{sec:package} provides a detailed description of the
\texttt{summclust} package which computes these variance estimators and
diagnostic measures.  The command uses the syntax:
\begin{small}
\begin{verbatim}
summclust varlist, cluster(varname) [options]
\end{verbatim}
\end{small}
\noindent The package has quite a few options and
can even be used by itself to estimate a linear regression model with
clustered disturbances. The last few sections of the paper illustrate
the use of \texttt{summclust} and provide evidence on the value of the
measures that it calculates. \Cref{sec:empirical} presents an
empirical illustration in which measures of leverage, partial
leverage, and influence are highly informative. \Cref{sec:examples}
discusses several special cases in which some or all of these measures
can be determined analytically. \Cref{sec:twoway} briefly discusses
two\tkk-way clustering, where \texttt{summclust} can be valuable even
though it is not explicitly designed to handle this case.
\Cref{sec:sims} describes some simulation experiments which suggest
that it may be desirable to report many of the quantities calculated
by \texttt{summclust}, and \Cref{sec:conc} concludes.

\section{Clustering, Leverage, Influence, and the Jackknife}
\label{sec:everything}

We focus on the linear regression model
\begin{equation}
\label{eq:lrmodel} 
\biy_g =\biX_g\bbeta + \biu_g, \quad g=1,\ldots,G,
\end{equation}
where the data have been divided into $G$ disjoint clusters. The
\th{g} cluster has $N_g$ observations, so that the sample size is $N =
\sum_{g=1}^G N_g$. In \eqref{eq:lrmodel}, $\biX_g$ is an $N_g\times k$
matrix of regressors, $\bbeta$ is a $k$-vector of coefficients,
$\biy_g$ is an $N_g$-vector of observations on the regressand, and
$\biu_g$ is an $N_g$-vector of disturbances (or error terms). The
$\biX_g$ may of course be stacked into an $N\times k$ matrix
$\biX$\tn, and likewise the $\biy_g$ and $\biu_g$ may be stacked into
$N$-vectors $\biy$ and $\biu$, so that \eqref{eq:lrmodel} can be
rewritten as $\biy = \biX\tn\bbeta + \biu$.

Dividing the sample into clusters only becomes meaningful if we make
assumptions about the disturbance vectors $\biu_g$ and, consequently,
the score vectors $\bis_g = \biX_g^\top \biu_g$. For a correctly
specified model, $\E(\bis_g)=\bzero$ for all $g$. We further assume
that
\begin{equation}
\label{eq:Sigma_g}
\E(\bis_g\bis_g^\top) = \bSigma_g \quad\mbox{and}\quad
\E(\bis_g\bis_{g'}^\top) = \bzero, \quad g,g'=1,\ldots,G,\quad g'\ne g,
\end{equation}
where $\bSigma_g$ is the symmetric, positive semidefinite variance
matrix of the scores for the \th{g} cluster. The second assumption in
\eqref{eq:Sigma_g} is crucial. It says that the scores for every
cluster are uncorrelated with the scores for every other cluster. We
take the number of clusters $G$ and the allocation of observations to
clusters as given. The important issue of how to choose the clustering
structure, perhaps by testing for the correct level of clustering, is
discussed in detail in \citet*{MNW-testing}.

The OLS estimator of $\bbeta$ is
\begin{equation*}
\hat\bbeta = (\biX^\top\!\biX)^{-1}\biX^\top\biy
= \bbeta_0 +  (\biX^\top\!\biX)^{-1}\biX^\top\biu,
\end{equation*}
where the second equality depends on the assumption that the data are
actually generated by \eqref{eq:lrmodel} with true value $\bbeta_0$. It 
follows that
\begin{equation}
\hat\bbeta - \bbeta_0 = (\biX^\top\!\biX)^{-1}\sum_{g=1}^G \biX_g^\top\biu_g
= \Big(\tn\sum_{g=1}^G\biX_g^\top\!\biX_g\Big)^{\!\!-1} \sum_{g=1}^G \bis_g.
\label{eq:betahat}
\end{equation}
From the rightmost expression in \eqref{eq:betahat}, we see that the
distribution of $\hat\bbeta$ depends on the disturbance subvectors
$\biu_g$ only through the distribution of the score vectors $\bis_g$.
Asymptotic inference commonly uses the empirical score vectors
$\hat\bis_g = \biX_g^\top \hat\biu_g$, in which the $\biu_g$ are
replaced by the residual subvectors $\hat\biu_g$, to estimate the
variance matrix of the $\bis_g$. This should work well if the sum of
the $\bis_g$, suitably normalized, is well approximated by a
multivariate normal distribution with mean zero, and if the $\bis_g$
are well approximated by the $\hat\bis_g$. However, asymptotic
inference can be misleading when either of these approximations is
poor.


It follows immediately from \eqref{eq:betahat} that an estimator of the 
variance of $\hat\bbeta$ may be based on the usual sandwich formula,
\begin{equation}
\label{eq:trueV}
(\biX^\top\!\biX)^{-1} \Big(\tn\sum_{g=1}^G \bSigma_g\tn\Big) 
(\biX^\top\!\biX)^{-1}.
\end{equation}
The natural way to estimate \eqref{eq:trueV} is to replace the
$\bSigma_g$ matrices by their empirical counterparts, that is, the 
$\hat\bis_g\hat\bis_g^\top$. If, in addition, we multiply by a
correction for degrees of freedom, we obtain the cluster-robust
variance estimator, or CRVE,
\begin{equation}
\mbox{CV$_{\tn1}$:}\qquad
\frac{G(N-1)}{(G-1)(N-k)}
(\biX^\top\!\biX)^{-1}
\Big(\tn\sum_{g=1}^G \hat\bis_g\hat\bis_g^\top\Big) (\biX^\top\!\biX)^{-1}.
\label{eq:CV1}
\end{equation}
This is by far the most widely used CRVE in practice, and it is the
default one implemented in \texttt{Stata}; alternatives to this
estimator will be discussed in \Cref{sec:jack}. When $G=N$\tn, the
CV$_{\tn1}$ estimator reduces to the familiar HC$_1$ estimator
\citep*{MW_1985} that is robust only to heteroskedasticity of unknown
form.



The fundamental unit of inference for clustered observations is not
the observation but the cluster; this is evident from
\eqref{eq:betahat}, \eqref{eq:trueV}, and \eqref{eq:CV1}. The
asymptotic theory for cluster-robust inference has been analyzed by
\citet*{DMN_2019} and \citet{HansenLee_2019} under the assumption that
$G\to\infty$. The quality of the asymptotic approximation depends on
the number of clusters $G$ and the heterogeneity of the score vectors
\citep*{MNW-guide}. The more the distributions of the scores vary
across clusters, the worse the asymptotic approximation will likely
be. Heterogeneity can arise from variation in cluster sizes and/or
from variation in the distributions of the disturbances, the
regressors, or both. As we discuss in
\Cref{sec:influential,sec:report,sec:sims}, leverage, partial
leverage, and summary statistics based on them provide useful measures
of heterogeneity across clusters.

Inference about $\bbeta$ is typically based on cluster-robust
$t$-statistics and Wald statistics. If $\beta_j$ denotes the \th{j}
element of $\bbeta$ and $\beta_{0j}$ is its value under the null
hypothesis, then the appropriate $t$-statistic is
\begin{equation*}
t_j = \frac{\hat\beta_j - \beta_{0j}}{\textrm{s.e.}(\hat\beta_j)},
\end{equation*}
where $\hat\beta_j$ is the OLS estimate, and
$\textrm{s.e.}(\hat\beta_j)$ is the square root of the \th{j} diagonal
element of \eqref{eq:CV1}. Under extremely strong assumptions
\citep*{BCH_2011}, it can be shown that $t_j$ asymptotically follows
the $t(G-1)$ distribution. Conventional inference in \texttt{Stata}
and other programs is based on this distribution.

As the articles cited in the second paragraph of \Cref{sec:intro}
discuss, inference based on $t_j$ and the $t(G-1)$ distribution can be
unreliable when $G$ is small and/or the clusters are severely
heterogeneous. This is true to an even greater extent for Wald tests
of two or more restrictions \citep*{PT_2018}. The measures of leverage
and partial leverage at the cluster level that we introduce in the
next section may help to identify the sort of heterogeneity that is
likely to make inference unreliable.


Instead of using the $t(G-1)$ distribution, we can obtain both $P$
values for $t_j$ and confidence intervals for $\beta_j$ by employing
the wild cluster restricted (or WCR) bootstrap \citep*{CGM_2008}. It
can sometimes provide much more reliable inferences than the
conventional approach; see \Cref{sec:sims}. \citet*{RMNW} describes a
computationally efficient implementation of this method in the
\texttt{Stata} package \texttt{boottest}. \citet*{MNW-bootknife}
proposes new versions of the wild cluster bootstrap that involve
transforming the empirical scores. When $G$ is reasonably large and
the clusters are not very heterogeneous, inferences based on the WCR
bootstrap and inferences based on CV$_{\tn1}$ $t$-statistics combined
with the $t(G-1)$ distribution will often be very similar. When they
differ noticeably, neither should be relied upon without further
investigation.


\Cref{sec:jack} discusses two CRVEs, which we refer to as CV$_{\tn3}$
and CV$_{\tn3{\rm J}}$, that are both based on the cluster jackknife.
In practice, these estimators are often extremely similar. CV$_{\tn3}$
and CV$_{\tn3{\rm J}}$ tend to yield more reliable inferences in
finite samples than does CV$_{\tn1}$, especially when the clusters are
quite heterogeneous; see \Cref{sec:sims} and \citet*{MNW-bootknife}.
Based on this simulation evidence, we recommend computing either
CV$_{\tn3}$ or CV$_{\tn3{\rm J}}$ essentially all the time. This is
easy to do using \texttt{summclust}.

\subsection{Identifying High-Leverage and Influential Clusters}
\label{sec:influential}

At the observation level, there are three classic measures of
heterogeneity, namely, leverage, partial leverage, and influence
\citep*{BKW_1980,CH_1986}. In this section, we propose analogous
measures at the cluster level.

Measures of leverage at the observation level are based on how much
the residual for observation $i$ changes when that observation is
omitted from the regression. If $h_i$ denotes the \th{i} diagonal
element of the ``hat matrix'' $\biH = \biP_\biX =
\biX(\biX^\top\!\biX)^{-1}\biX^\top$\tn, then omitting the \th{i}
observation changes the \th{i} residual from $\hat u_i$ to $\hat
u_i/(1-h_i)$. Because $0 < h_i < 1$, this delete\tkk-one residual is
always larger in absolute value than $\hat u_i$. The factor by which
the delete\tkk-one residual exceeds $\hat u_i$ increases with $h_i$.
Since the average of the $h_i$ is $k/N$\tn, observations with values
of $h_i$ substantially larger than $k/N$ may reasonably be said to
have high leverage.




Dropping the \th{g} cluster when we estimate $\bbeta$ yields the
delete\tkk-one\tkk-cluster estimate $\hat\bbeta^{(g)}$\tn. Using
$\hat\bbeta^{(g)}$ in place of $\hat\bbeta$ changes the residual
vector for the \th{g} cluster from $\hat\biu_g$ to $\hat\biu_g^{(g)}$.
These delete\tkk-one\tkk-cluster residual vectors can be written in 
two ways:
\begin{equation*}
\hat\biu_g^{(g)} = \biy_g - \biX_g \hat{\bbeta}^{(g)}
= (\bfI -\biH_g)^{-1}\hat\biu_g. 
\end{equation*}
In the rightmost expression above,
\begin{equation*}
\biH_g = \biX_g(\biX^\top\!\biX)^{-1}\biX_g^\top
\end{equation*}
is the $N_g\times N_g$ diagonal block of $\biH$ that corresponds to
cluster~$g$. The matrix $\biH_g$ is the cluster analog of the scalar
$h_i$. Of course, it is not feasible to report the $\biH_g$. In fact,
when any of the clusters is sufficiently large, even computing and
storing these matrices may be challenging. As a measure of leverage,
we therefore suggest using a matrix norm of the $\biH_g$.
Specifically, we suggest the scalar
\begin{equation}
\label{eq:traceXX}
L_g = \Tr(\biH_g) =
\Tr\!\big(\biX_g^\top\!\biX_g(\biX^\top\!\biX)^{-1}\big).
\end{equation}
When the \th{g} cluster contains just one observation, say the \th{i},
then $L_g = h_i$. Thus, in this special case, the leverage
measure that we are proposing reduces to the usual measure of leverage
at the observation level.


The trace in \eqref{eq:traceXX} is the nuclear norm of the matrix
$\biH_g$. In general, the nuclear norm of a matrix $\biA$ is the sum
of the singular values of $\biA$. When $\biA$ is symmetric and
positive semidefinite, the singular values are equal to the
eigenvalues, which are non-negative. Since the trace of any square
matrix is equal to the sum of the eigenvalues, the trace of a
symmetric and positive semidefinite matrix is also its nuclear norm.
In principle, we could report any norm of the $\biH_g$ matrices, but
the nuclear norm is particularly easy to compute. Also, because it is
linear, we can sum over $g$ and take the sum inside the norm just as
if the $\biH_g$ were scalars. Since $\sum_{g=1}^G \biX_g^\top\!\biX_g
= \biX^\top\!\biX$, this result means that $G^{-1} \sum_{g=1}^G
\Tr(\biH_g) = k/G$, which is analogous to the result that the average
of the $h_i$ over all observations is $k/N$\tn.

High-leverage clusters can be identified by comparing the $L_g$ to
$k/G$, their average. If, for some cluster $h$, $L_h$ is substantially
larger than $k/G$, then cluster $h$ may be said to have high leverage.
Just how much larger $L_h$ has to be is a matter of judgement. A
cluster with $L_h=2k/G$ probably does not qualify, but a cluster with
$L_h=5k/G$ probably does. Cluster~$h$ can have high leverage either
because $N_h$ is considerably larger than $G/N$ or because the matrix
$\biX_h$ is somehow extreme relative to the other $\biX_g$ matrices,
or both. We can compare the leverage of any two clusters by forming
ratios. For example, if $L_1 = 3$ and $L_2 = 1$, then we can say that
the first cluster has three times the leverage of the second cluster.


The leverage measure we suggest in \eqref{eq:traceXX} shows the
potential impact of a specified cluster on residuals and fitted
values, but not on any particular regression coefficient. When
interest focuses on just one such coefficient, say the \th{j}\tn, it
may be more interesting to calculate the partial leverage of each
cluster. The concept of partial leverage was introduced, for
individual observations, in \citet*{CW_1980}. Let
\begin{equation*}
\acute\bix_j = \big(\bfI -
\biX_{[j]}\big(\biX_{[j]}^\top\biX_{[j]}\big)^{\tn-1}
\biX_{[j]}^\top\big)\bix_j,
\end{equation*}
where $\bix_j$ is the vector of observations on the \th{j} regressor,
and $\biX_{[j]}$ is the matrix of observations on all the other
regressors. Thus $\acute\bix_j$ denotes $\bix_j$ after all the other
regressors have been partialed out. The partial leverage of
observation $i$ is simply the \th{i} diagonal element of the
matrix
$\acute\bix_j(\acute\bix_j^\top\acute\bix_j)^{-1}\acute\bix_j^\top$,
which is just $\acute x_{ji}^2/(\acute\bix_j^\top\acute\bix_j)$, where
$\acute x_{ji}^2$ is the \th{i} element of $\acute\bix_j$.

The analogous measure of partial leverage for cluster $g$ is
\begin{equation}
L_{gj} = \frac{\acute\bix_{gj}^\top\acute\bix_{gj}}
{\acute\bix_j^\top\acute\bix_j}\tk,
\label{eq:partial}
\end{equation}
where $\acute\bix_{gj}$ is the subvector of $\acute\bix_j$
corresponding to the \th{g} cluster. This is what \eqref{eq:traceXX}
reduces to if we replace $\biX$ and $\biX_g$ by $\acute\bix_j$ and
$\acute\bix_{gj}$, respectively. It is easy to calculate the partial
leverage for every cluster for any coefficient of interest. The
average of the $L_{gj}$ is evidently $1/G$, so that if cluster $h$ has
$L_{hj} >\!> 1/G$, it has high partial leverage for the \th{j}
coefficient. Moreover, as we will see in \Cref{sec:sims}, the
empirical distribution of the $L_{gj}$ across clusters seems to
provide useful diagnostic information.

%





\citet{AY-IV} derives a measure of cluster-level leverage for the
first-stage regression used to obtain a linear instrumental variables
estimator. The paper calls $L_{gj}$ the group~$g$ ``share of 
coefficient leverage'' for instrument~$j$ and then uses the maximum of
the $L_{gj}$ over all the instruments excluded from the structural
equation as a measure of the leverage of cluster~$g$. Using
simulations based on 1309 IV regressions from 30 published papers, the
paper finds that inference is much less reliable for models where one
or two clusters have high leverage in the first-stage regression than
for models where no clusters do so.


One possible consequence of heterogeneity is that the estimates may
change a lot when certain clusters are deleted. It can therefore be
illuminating to delete one cluster at a time, so as to see how
influential each cluster is. To do this in a computationally efficient
manner, \texttt{summclust} first computes the cluster-level matrices
and vectors
\begin{equation}
\biX_g^\top\!\biX_g \quad\mbox{and}\quad \biX_g^\top\biy_g, \quad
g=1,\ldots,G.
\label{eq:subthings}
\end{equation}
These are then used to construct $\biX^\top\!\biX$ and
$\biX^\top\biy$, and the vector of least squares estimates when
cluster $g$ is deleted is computed as
\begin{equation}
\label{eq:delone}
\hat\bbeta^{(g)} = (\biX^\top\!\biX - \biX_g^\top\!\biX_g)^{-1}
(\biX^\top\biy - \biX_g^\top\biy_g).
\end{equation}
Unless $k$ is extremely large, it should generally not be expensive to
compute $\hat\bbeta^{(g)}$ for every cluster using \eqref{eq:delone}.
\texttt{summclust} simply has to invert $G$ matrices, each of them
$k\times k$, and then multiply each of those matrices by a $k$-vector.

Especially when they vary a lot, the $\hat\bbeta^{(g)}$ can reveal a
great deal about the sample. In addition, as we shall see in
\Cref{sec:jack}, they may be used to calculate jackknife variance
matrices. When there is a parameter of particular interest,
say~$\beta_j$, it may be a good idea to report the $\hat\beta_j^{(g)}$
for $g=1,\ldots,G$ in either a histogram or a table. By default, 
\texttt{summclust} creates several figures with these and other
cluster-level statistics. If $\hat\beta_j^{(h)}$ differs greatly from
$\hat\beta_j$ for some cluster $h$\tn, then cluster $h$ is evidently
influential.


In a few extreme cases, there may be a cluster $h$ for which it is
impossible to compute $\hat\beta_j^{(h)}$\tn. This will happen, for
example, when the regressor corresponding to $\beta_j$ is a treatment
dummy and cluster $h$ is the only treated one. This is an extreme
example of the problem of few treated clusters, and inferences based
on either the $t(G-1)$ distribution or the WCR bootstrap are likely to
be completely unreliable in this case \citep*{MW-TPM,MW-EJ,MW-RI}.


Identifying influential clusters by comparing the $\hat\bbeta^{(g)}$
with $\hat\bbeta$ is very similar to identifying influential
observations using the classic methods discussed in \citet*{BKW_1980}
and \citet*{CH_1986}; for an interesting recent extension, see
\citet*{BGM_2021}. Unlike the leverage measures, the
$\hat\beta^{(g)}_j$ may be either positive or negative, must depend on
the $\biy_g$, and necessarily vary across clusters. They may sometimes
reveal features of the model or dataset that require further
investigation. Perhaps the model does not seem to apply to some
clusters, or perhaps there are measurement errors or observations that
have been miscoded.

Regression models often include cluster fixed effects. When one of the
regressors is a fixed-effect dummy for cluster $g$, the matrices
$\biX_g^\top\!\biX_g$ and $\biX^\top\!\biX - \biX_g^\top\!\biX_g$ are
singular. This will cause the calculation in \eqref{eq:delone} to fail
unless a generalized inverse routine, such as the \texttt{invsym}
routine in \texttt{Mata}, is used. Although \texttt{summclust} uses
this routine, it also provides options to avoid the problem, and save
some computer time, by partialing out the fixed-effect dummies prior
to computing the cluster-level matrices and vectors in
\eqref{eq:subthings}; see \Cref{sec:package}.

Partialing out cluster fixed effects means replacing $\biX$ and $\biy$
by $\tilde\biX$ and $\tilde\biy$, the deviations from their cluster
means. For example, the element of $\tilde\biy$ corresponding to the
\th{j} observation in the \th{g} cluster is $y_{g,j} -
N_g^{-1}\sum_{i=1}^{N_g} y_{g,i}$. The \th{g} subvector of
$\tilde\biy$ is $\tilde\biy_g$, and the \th{g} submatrix of
$\tilde\biX$ is $\tilde\biX_g$. Since there is just one fixed effect
per cluster, $\tilde\biy_g$ depends solely on $\biy_g$, and
$\tilde\biX_g$ depends solely on $\biX_g$. The calculations in
\eqref{eq:traceXX} and \eqref{eq:delone} are now based on
$\tilde\biX^\top\!\tilde\biX$\tn, $\tilde\biX^\top\tilde\biy$, the
$\tilde\biX_g^\top\!\tilde\biX_g$, and the
$\tilde\biX_g^\top\tilde\biy_g$. Importantly, the sum of the $L_g$ is
now equal to the number of columns in $\tilde\biX$ instead of the
number of columns in $\biX$\tn.


\subsection{Two Jackknife Variance Matrix Estimators}
\label{sec:jack}


Although the CV$_{\tn1}$ variance estimator defined in \eqref{eq:CV1}
is very widely used, it often does not have good finite\tkk-sample
properties. Two alternative CRVEs, which are usually known as
CV$_{\tn2}$ and CV$_{\tn3}$, were proposed in \citet*{BM_2002}. They
are the cluster analogs of the hetero\-skedasticity-consistent
estimators HC$_2$ and HC$_3$, which are appropriate when the $u_i$ are
independent. These names were coined in \citet*{MW_1985}, which
proposed HC$_3$ as a jackknife variance estimator. In the remainder of
this section, we focus on CV$_{\tn3}$, because CV$_{\tn2}$ is not a
jackknife estimator and is not amenable to the computational methods
that we propose; for more on it, see \citet*{Imbens_2016},
\citet*{PT_2018}, and \citet*{NAAMW_2020}. \texttt{Stata 18} added the
ability to rapidly calculate CV$_{\tn2}$ standard errors, using the
option \texttt{vce(hc2 clustvar)}.  Simulations in
\citet*{MNW-bootknife} suggest that CV$_{\tn2}$ is preferred to
CV$_{\tn1}$, but that CV$_{\tn3}$ is almost always preferred to
CV$_{\tn2}$.






CV$_{\tn3}$ can be written in several ways. One of them is
\begin{equation}
\mbox{CV$_{\tn3}$:}\qquad \frac{G-1}{G}
(\biX^\top\biX)^{-1}\Big(\tk\sum_{g=1}^G \ddot\bis_g\ddot\bis_g^\top
\Big)(\biX^\top\biX)^{-1},
\label{eq:CV3}
\end{equation}
where the modified score vectors $\ddot\bis_g$ are defined as
\begin{equation*}
\ddot\bis_g = \biX_g^\top\biM_{gg}^{-1}\tk\hat\biu_g.
\end{equation*}
Here $\biM_{gg} = \bfI_{N_g} -
\biX_g(\biX^\top\biX)^{-1}\tn\biX_g^\top$ is the diagonal block
corresponding to the \th{g} cluster of the projection matrix
$\biM_\biX$, which satisfies $\hat\biu = \biM_\biX\biu$. Although
computing CV$_{\tn3}$ using \eqref{eq:CV3} works well when all the
$N_g$ are very small, it becomes expensive, or perhaps computationally
infeasible, when one or more of the $N_g$ is large. The problem is
that an $N_g\times N_g$ matrix needs to be stored and inverted for
every cluster. \citet*{NAAMW_2020} proposes a method that is much
faster for large clusters, versions of which apply to both CV$_{\tn2}$
and CV$_{\tn3}$. However, recognizing that CV$_{\tn3}$ is a jackknife
estimator makes a method that is even simpler and usually faster
available.


There are actually two cluster jackknife estimators of
$\var(\hat\bbeta)$. The simplest is probably
\begin{equation}
\mbox{CV$_{\tn3{\rm J}}$:}\qquad
\frac{G-1}{G} \sum_{g=1}^G (\hat\bbeta^{(g)} -
\bar\bbeta)(\hat\bbeta^{(g)} - \bar\bbeta)^\top,
\label{eq:jack}
\end{equation}
where $\bar\bbeta$ is the sample mean of the $\hat\bbeta^{(g)}$, which
were defined in \eqref{eq:delone}. The expression in \eqref{eq:jack}
is the cluster analog of the usual jackknife variance matrix estimator
given in \citet*[eqn.\ (11)]{MW_1985}. Each of the $\hat\bbeta^{(g)}$
is obtained by deleting a cluster instead of an observation, and the
summation is over clusters instead of observations. If $\bar\bbeta$ in
\eqref{eq:jack} is replaced by $\hat\bbeta$, we obtain instead
\begin{equation}
\mbox{CV$_{\tn3}$:}\qquad
\frac{G-1}{G} \sum_{g=1}^G (\hat\bbeta^{(g)} - \hat\bbeta)
(\hat\bbeta^{(g)} - \hat\bbeta)^\top.
\label{eq:jack3}
\end{equation}
This version of CV$_{\tn3}$ is numerically identical to the one in 
\eqref{eq:CV3} \citep*[Section~3]{MNW-bootknife}. Unless all the 
clusters are very small, computing CV$_{\tn3}$ using \eqref {eq:jack3} 
is much faster than using \eqref{eq:CV3}; timings are reported in 
\citet*{MNW-bootknife}.




Many discussions of jackknife variance estimation follow
\citet*{Efron_79} and use $\bar\bbeta$ as in \eqref{eq:jack}, but
others, including \citet*{BM_2002}, use $\hat\bbeta$ as in
\eqref{eq:jack3}. Although these two jackknife variance estimators are
asymptotically the same, they are rarely equal, since CV$_{\tn3}$
minus CV$_{\tn3{\rm J}}$ is a positive semi-definite matrix. In
practice, however, they tend to be very similar
\citep*{MNW-bootknife}, and there seems to be no good reason to expect
either CV$_{\tn3}$ or CV$_{\tn3{\rm{J}}}$ to perform better in
general. Interestingly, the original HC$_3$ estimator proposed in
\citet*{MW_1985} is actually the analog of CV$_{\tn3{\rm J}}$. The
modern version of HC$_3$, which is the analog of CV$_{\tn3}$, seems to
be due to \citet*[Chapter~16]{DM_1993}. This version of HC$_3$ is
normally computed by dividing each residual by the corresponding
diagonal element of $\biM_\biX$, and the factor of $(N-1)/N$ is
usually (but incorrectly) omitted.


The factor of $(G-1)/G$ in both \eqref{eq:jack} and \eqref{eq:jack3}
is designed to compensate for the tendency of the $\hat\bbeta^{(g)}$
to be too spread out. This factor is the analog of the usual factor of
$(N-1)/N$ for a jackknife variance matrix at the individual level. It
implicitly assumes that all clusters are the same size and perfectly
balanced, with disturbances that are independent and homoskedastic. In
this special case, the estimators CV$_{\tn3}$ and CV$_{\tn3{\rm{J}}}$
would be identical and unbiased \citep*{BM_2002}. These estimators
are already available in \texttt{Stata}. When used with the
\texttt{cluster} option, the \texttt{vce(jackknife)} option computes
CV$_{\tn3{\rm{J}}}$ standard errors, and the
\texttt{vce(jackknife,mse)} option computes CV$_{\tn3}$ standard
errors. Because it is specialized for linear regression models, the
implementation in \texttt{summclust} is quite a bit faster.


Both jackknife estimators may readily be used to compute
cluster-robust $t$-statistics. Because there are $G$ terms in the
summation, it seems natural to compare these with the $t(G-1)$
distribution, as usual. These procedures should almost always be more
conservative than $t$-tests based on the widely-used CV$_{\tn1}$
estimator. In an important recent paper, \citet*{Hansen-jack} shows
that CV$_{\tn3}$ has much better worst-case theoretical properties
than CV$_{\tn1}$. This strongly suggests that $t$-statistics based on
CV$_{\tn3}$ are likely to yield lower rejection frequencies than ones
based on CV$_{\tn1}$. The simulation results in \Cref{sec:sims} and in
\citet*{MNW-bootknife} are consistent with this conjecture.


When a model includes fixed effects, some care needs to be taken when
computing CV$_{\tn3}$ and CV$_{\tn3{\rm J}}$. As noted in
\Cref{sec:influential}, it is computationally attractive to partial
out fixed effects prior to calculating $\hat\bbeta$. However, if we
were to partial out any arbitrary regressors prior to computing the
delete\tkk-one\tkk-cluster estimates in \eqref{eq:delone}, then the
computed $\hat\bbeta^{(g)}$ would depend on the values of the
partialed-out regressors for the full sample, including those in the
\th{g} cluster. Consequently, the values of CV$_{\tn3}$ and
CV$_{\tn3{\rm J}}$ will be incorrect if we partial out any regressor
that affects more than one cluster (such as industry-level fixed
effects with firm-level clustering). The regressors that are partialed
out must be cluster fixed effects or fixed effects at a finer level
(such as firm-level fixed effects with industry-level clustering),
because each of them affects only one cluster. See the discussion of
the \texttt{absorb} and \texttt{fevar} options in \Cref{sec:package}.

It is possible that the vector $\bbeta$ is identified for the full 
sample but not when one cluster is deleted. For example, consider the 
coefficient on a dummy variable that takes on non-zero values only for
observations in the \th{g} cluster. This coefficient cannot be 
identified when cluster $g$ is omitted. In such a case, the matrix 
$\biX^\top\!\biX - \biX_g^\top\!\biX_g$ in \eqref{eq:delone} is 
singular, and CV$_{\tn3}$ and CV$_{\tn3{\rm J}}$ cannot be computed
using an ordinary matrix inverse. However, because \texttt{summclust}
uses the \texttt{invsym} function in \texttt{Stata}, which implements
a generalized inverse, the offending element of $\hat\bbeta^{(g)}$ is
simply replaced by~$0$. The package therefore checks whether any of
the $\hat\beta^{(g)}$ coefficients of interest are equal to~$0$ and
issues a warning when they are; see \Cref{sec:package}.






There may be more than one set of fixed effect that are invariant at
the cluster level. For example, imagine an analysis of students' test
scores where the researcher wants to control for both school and
neighborhood fixed effects and cluster the standard errors at the
state level. In this case, neither of \texttt{Stata's} built-in
\texttt{regress} and \texttt{areg} commands can produce an estimate of
CV$_{\tn3}$, because the fixed effects for schools and neighborhoods
in state $g$ cannot be identified when state~$g$ is omitted. However,
\texttt{summclust} can produce such an estimate.





\subsection{What Should Be Reported}
\label{sec:report}

We believe that investigators should routinely compute the $L_g$. They
should also compute the $L_{gj}$ for any coefficient(s) of particular
interest. In some cases, the $L_g$ and the $L_{gj}$ will be roughly
proportional to the $N_g$ (the cluster sizes). That in itself would be
informative. It may be even more interesting, however, to find that
the relative size of $L_g$ and/or $L_{gj}$ for some cluster(s) $g$ is
much larger, or much smaller, than the relative size of $N_g$.


When the number of clusters is small, it is easy enough to look at all
the $N_g$, $\hat\beta^{(g)}_j$, $L_g$, and $L_{gj}$ to see whether any
clusters are unusually large, unusually influential, or have unusually
high leverage or partial leverage. Once $G$ exceeds 10 or 12, however,
it may be more informative to report summary statistics or to plot
these quantities. The \texttt{summclust} package always reports the
minimum, first quartile, median, mean, third quartile, and maximum of
the $N_g$ and the~$L_g$. It also reports these quantities for the
$L_{gj}$ and the $\hat\beta^{(g)}_j$ for the specified regressor~$j$,
and by default it provides a figure containing four scatterplots of
the $L_g$ and the $L_{gj}$ against the $N_g$ and the~$\hat\beta^{(j)}$;
see \Cref{sec:package,sec:empirical}.

Another possibility is to report a few summary statistics, as
\texttt{summclust} also does. Consider a generic (positive) quantity
$a_g$, which might denote any of $N_g$, $L_g$, or $L_{gj}$ for
$g=1,\ldots,G$. It seems plausible that inference may be unreliable
when any of the $a_g$ vary substantially across clusters, and we
provide some evidence to support this conjecture in \Cref{sec:sims}.

There are many measures of how much the distribution of the $a_g$
differs from what it would be in the perfectly balanced case. One
of these is the scaled variance
\begin{equation}
V_s(a_{\bullet}) = \frac{1}{(G-1)\bar a^2} \sum_{g=1}^G (a_g - \bar a)^2,
\label{eq:coefvar}
\end{equation}
where the argument $a_{\bullet}$ is to be interpreted as the entire
set of $a_g$ for $g=1,\ldots,G$, and $\bar a$ denotes the arithmetic
mean, which is $N/G$ for the $N_g$, $k/G$ for the $L_g$, and $1/G$ for
the $L_{gj}$. These are all positive numbers, so it is reasonable to
scale by their squares. Larger values of $V_s$ imply that the $a_g$
are more variable across clusters, relative to their mean. We could
report either $V_s$ or its square root, which is often called the
coefficient of variation. In the perfectly balanced case, $V_s=0$. By
default, \texttt{summclust} reports the coefficient of variation for
the cluster sizes, the leverages, the partial leverages, and the
$\hat\beta_j^{(g)}$\tn.


Another possibility, which is only valid for positive quantities, is
to report one or more alternative sample means. The more these differ
from the arithmetic mean, the more heterogeneous must be the clusters.
Three common alternatives to the arithmetic mean are the harmonic,
geometric, and quadratic means:
\begin{equation*}
\bar a_{\rm harm} = \left(\frac{1}{G}\sum_{g=1}^G
1/a_g\right)^{\!\!\!-1}\tn,
\quad \bar a_{\rm geo} = \left(\prod_{g=1}^G a_g\right)^{\!\!\!1/G}\tn,
\;\;\mbox{and}\;\;
\bar a_{\rm quad} = \left(\frac{1}{G}\sum_{g=1}^G
a_g^2\right)^{\!\!\!1/2}\tn.
\end{equation*}
Unless all the $a_g$ are the same, the harmonic and geometric means
will be less than the arithmetic mean $\bar a$, and the quadratic mean
(which has the same form as the root mean squared error of an
estimator) will be greater than~$\bar a$. \texttt{summclust}
optionally reports all three of these alternative means, along with
the ratio of each of them to $\bar a$. The three ratios provide
scale\tkk-free measures of cluster heterogeneity; the closer they are
to one, the more homogeneous are the clusters.

Another way to quantify the heterogeneity of the cluster sizes and the
regressors is to calculate $G^*$\tn, the ``effective number of
clusters,'' as proposed in \citet*{CSS_2017}. The value of $G^*$ 
depends on the coefficient $j$ for which it is being computed and on 
a parameter $\rho$ to be discussed below, so we denote it
$G^*_j(\rho)$. It is defined as
\begin{equation}
G^*_j(\rho) = \frac{G}{1 + \Gamma_j(\rho)}\tk, \quad
\Gamma_j(\rho) = \frac1G \sum_{g=1}^G \Big(\frac{\gamma_{gj}(\rho) -
\bar\gamma_j(\rho)}{\bar\gamma_j(\rho)}\Big)^{\!2}, \quad \bar\gamma_j(\rho)
= \frac1G\sum_{g=1}^G \gamma_{gj}(\rho),
\label{eq:gstar}
\end{equation}
where $0\le\rho\le1$, and the $\gamma_{gj}(\rho)$ are given by
\begin{equation}
\gamma_{gj}(\rho) = \bie_j^\top(\biX^\top\biX)^{-1}
\biX_g^\top\bOmega_g(\rho)\biX_g
(\biX^\top\biX)^{-1}\bie_j, \quad g=1,\ldots,G.
\label{eq:gammag}
\end{equation}
Here $\bie_j$ is a $k$-vector with 1 in the \th{j} position and 0
everywhere else, so that $\bie_j^\top(\biX^\top\biX)^{-1}$ is the
\th{j} row of $(\biX^\top\biX)^{-1}$\tn, and $\bOmega_g(\rho)$ is an
$N_g\times N_g$ matrix with 1 on the principal diagonal and $\rho$
everywhere else. It is easy to see that
\begin{equation}
\bOmega_g(\rho) = \rho\tk\biota\biota^\top + (1-\rho)\tkk\bfI,
\label{eq:omega}
\end{equation}
where $\biota$ is an $N_g$-vector of 1s, and $\bfI$ is an $N_g\times
N_g$ identity matrix. Notice that $\Gamma_j(\rho)$ is just the scaled
variance of the $\gamma_{gj}(\rho)$; compare \eqref{eq:coefvar}.


The parameter $\rho$ may be interpreted as the intra-cluster 
correlation coefficient for a model with cluster-level random effects.
Since $\rho$ is unknown, \citet*{CSS_2017} suggests calculating
$G^*_j(1)$ as a sort of worst case. However, when there are
cluster-level fixed effects, or fixed effects at a finer level nested
within clusters, they will absorb all of the intra-cluster
correlation. Thus it does not make sense to specify $\rho>0$ in either
of these cases. It does seem natural to use $G^*_j(0)$, however,
because the amount of intra-cluster correlation that remains in models
with cluster fixed effects is often quite small.

From \eqref{eq:gammag} and \eqref{eq:omega}, we see that
\begin{equation}
\biX_g^\top\bOmega_g(\rho)\biX_g = \rho (\biota^\top\!\biX_g)^\top
(\biota^\top\!\biX_g) + (1-\rho)\biX_g^\top\biX_g.
\label{eq:gammiddle}
\end{equation}
This result makes it inexpensive to compute the $\gamma_{gj}(\rho)$ for
any value of $\rho$ by first computing them for $\rho=0$ and $\rho=1$.
The needed equations are
\begin{equation}
\begin{aligned}
\gamma_{gj}(0) &= \biw_j^\top\biX_g^\top\biX_g\tk\biw_j,\\
\gamma_{gj}(1) &=
(\biota^\top\!\biX_g\tk\biw_j)^\top(\biota^\top\!\biX_g\tk\biw_j),
\;\mbox{and}\\
\gamma_{gj}(\rho) &= \rho\tk\gamma_{gj}(1) + (1-\rho)\gamma_{gj}(0),
\end{aligned}
\label{eq:gammarho}
\end{equation}
where $\biw_j$ is the \th{j} column of $(\biX^\top\biX)^{-1}$. After
obtaining the $\gamma_{gj}(\rho)$ from \eqref{eq:gammarho}, it is trivial
to compute $G^*_j(\rho)$ using \eqref{eq:gstar}. Evidently,
$G^*_j(\rho)$ is always less than $G$. When it is much smaller
than~$G$, it can provide a useful warning.


Suppose that we have partialed out cluster fixed effects prior to
computing $G^*_j(\rho)$. Then the first term on the right-hand side of
\eqref{eq:gammiddle} should in theory be a zero matrix, because every
column of $\biX_g$ should add to zero. In practice, however, the
limitations of floating-point arithmetic mean that this matrix will
actually contain extremely small positive numbers. This will cause the
computation of $G^*_j(\rho)$ to be numerically unstable. When the
fixed effects are not partialed out, similar but more complicated
numerical issues arise.

The \texttt{Stata} package \texttt{clusteff} discussed in
\citet*{LS_2018} is designed to calculate $G^*_j(\rho)$, with
$\rho=0.9999$ rather than $\rho=1$ by default to avoid numerical
instabilities. However, the only version of this package that we have
used does so in a computationally inefficient way that does not use
\eqref{eq:gammarho}. When any of the $N_g$ is large, it can take a
very long time, or even fail because \texttt{Stata} runs out of
memory. For example, it failed with some of the samples in 
\citet*{MNW-guide}.




Like $V_s(a_{\bullet})$ and the alternative sample means for measures
of leverage and partial leverage discussed above, $G^*_j(\rho)$ is
sensitive not only to variation in cluster sizes but also to other
features of the $\biX_g$ matrices. But it is not sensitive to
heteroskedasticity or to any other features of the disturbances.
\texttt{summclust} computes $G^*_j(0)$, $G^*_j(1)$, and (optionally)
$G^*_j(\rho)$ for a specified covariate. However, when there are
cluster fixed effects, or fixed effects nested within clusters, it
only computes $G^*_j(0)$. For example, it will not compute
$G^*_j(\rho)$ for $\rho\ne0$ whenever there are state\tkk-level fixed
effects and clustering at the region level.

The quantity $G^*_j(0)$ is very closely related to
$V_s(L_{{\bullet}j})$, where $L_{{\bullet}j}$ denotes the entire set
of $L_{gj}$, for $g=1,\ldots,G$. It is not hard to see that the
$\gamma_g(0)$ defined in \eqref{eq:gammag} and \eqref{eq:gammarho} are
equal to the $L_{gj}$ defined in \eqref{eq:partial} divided by
$\acute\bix_j^\top\acute\bix_j$. Since this makes the $\gamma_g(0)$
proportional to the $L_{gj}$, $V_s(L_{{\bullet}j})$ is numerically
identical to $\Gamma(0)$; compare \eqref{eq:coefvar} and the middle
equation in \eqref{eq:gstar}. Thus we see from the first equation in
\eqref{eq:gstar} that $G^*_j(0)$ is simply a monotonically decreasing
function of the scaled variance of our measures of partial leverage at
the cluster level. When $V_s(L_{{\bullet}j})$ is large, $G^*_j(0)$ is
necessarily much smaller than $G$.

\section{The \texttt{summclust} Package}
\label{sec:package}



The \texttt{summclust} package may be obtained from 
\texttt{SSC} or \url{https://github.com/mattdwebb/summclust}. It
implements the \texttt{summclust} command, which calculates a large
number of statistics to help assess cluster heterogeneity and also
provides CV$_{\tn3}$ and CV$_{\tn3{\rm J}}$ standard errors. The
package does not rely on any other \texttt{Stata} packages, but it
does require a version of \texttt{Stata} that provides \texttt{Mata}'s
\texttt{panelsum()} function (Version~13 or later).

We first present an overview of the \texttt{summclust} command,
followed by a simple illustration using the \texttt{webuse} dataset
\texttt{nlswork}.



\subsection{Syntax and options}

\noindent \textbf{\texttt{\underline{Syntax}}}


\begin{small}
\begin{verbatim}
summclust varlist, cluster(varname) [options]
\end{verbatim}
\end{small}

\noindent\textit{varlist:} the dependent variable, the independent
variable of interest, and other (binary or continuous) independent
variables. At least one additional regressor must be specified.
Time\tkk-series operators and factor variables are not permitted. 



\noindent\textit{cluster:} the clustering variable, for which the
number of unique values equals~$G$.







\begin{longtable}{@{}lY@{}}

\textbf{\textit{options}} & \textbf{Description} \\
\hline
\endhead

\texttt{fevar(varlist)} & creates fixed effects for each of the specified
variables, using \texttt{i.varname}.\\

\texttt{absorb(varname)} & partials out the variable \texttt{varname}
before computing other statistics. This option should only be used for
variables that are nested within the specified clusters. It can often
be computationally faster than using \texttt{fevar} and should be used
when there are cluster-level fixed effects in order to avoid singular
omit-one-cluster samples caused by those fixed effects. In cases with
an extremely large number of fixed effects, \texttt{summclust} may run
into memory issues. If so, one can use the \texttt{Stata} prefix
\texttt{jackknife} with the user-contributed command
\texttt{reghdfe}.\\
 

\texttt{\underline{jack}knife} & calculates the jackknife variance
estimator CV$_{\tn3{\rm J}}$ in addition to CV$_{\tn3}$.\\


\texttt{\underline{add}means} & displays the alternative sample means of 
the $N_g$, $L_g$, $L_{gj}$, and~$\hat\beta_j^{(g)}$\tn, as described in
\Cref{sec:report}. For the $N_g$, $L_g$, and~$L_{gj}$, it reports the 
harmonic, geometric, and quadratic means, as well as the ratio of each
of them to the arithmetic mean. For the~$\hat\beta_j^{(g)}$\tn, which can
be negative, only the quadratic mean and its ratio are reported,
because the harmonic and geometric means are not defined for negative
numbers.\\

\texttt{gstar} & calculates the effective number of clusters G*(0) and,
when there are no cluster (or subcluster) fixed effects, G*(1) as well.\\





\texttt{rho(scalar)} & calculates the effective number of clusters,
G*(rho), in addition to G*(0) and G*(1). This option can be used with
or without the \texttt{gstar} option. The value of rho must be between
0 and 1; the program ends with an error message when an invalid value
for rho is entered. If it is not valid to display~G*(rho), due to 
variables that are invariant within clusters, it reports that G*(rho) 
cannot be computed and displays only~G*(0). There is no reason to use
the \texttt{gstar} option when this option is used.\\


\texttt{\underline{tab}le} & displays the cluster-by-cluster values of
cluster size, leverage, partial leverage, and the
delete\tkk-one\tkk-cluster coefficient estimate. If $G > 52$, then 
the unformatted matrix is displayed instead of a table.\\

\texttt{\underline{sam}ple} & allows for sample restrictions. The
argument(s) for this option are whatever would follow the ``if'' in a
standard \texttt{regress} command. For instance, in order to restrict
the analysis to individuals 25 years of age or older based on a variable
 ``age'', \texttt{sample(age>=25)} should be added to the list of
 options.\\

\texttt{\underline{nog}raph} & suppresses creation of the figure, which
is otherwise shown by default.\\

\texttt{\underline{reg}table} & displays a full table of regression
output, similar to Stata's regress table, but with jackknife standard
errors. It reports CV$_{\tn3}$ standard errors by default, but it
instead reports CV$_{\tn3{\rm J}}$ standard errors when the
\texttt{jackknife} option is also specified. If $k > 52$, then the
unformatted matrix is displayed instead of a table.\\

\end{longtable}
\vskip -12pt
\hrule
\bigskip



\noindent\textbf{\texttt{\underline{Description}}}

\medskip

\noindent \textbf{\texttt{summclust}} is a stand-alone command for
summarizing cluster variability in several ways. It always calculates
measures of cluster-level influence and leverage, and it optionally
calculates the effective number of clusters. It also always reports
CV$_{\tn1}$ and CV$_{\tn3}$ standard errors for a single coefficient,
and it optionally reports a CV$_{\tn3{\rm J}}$ standard error as well.
If requested, it can calculate additional measures of cluster-level
heterogeneity. Unless it is told not to, it produces a figure which
can help identify the source of cluster level heterogeneity. Finally,
it can optionally produce a full table of regression results with
CV$_{\tn3}$ standard errors.



By default, \textbf{\texttt{summclust}} calculates the CV$_{\tn3}$
standard error based on \eqref{eq:CV3}. With well-behaved samples,
this should match the standard error calculated using either
\texttt{Stata}'s native \textbf{\texttt{jackknife:}}
\textbf{\texttt{reg y x, cluster(group)}} or \textbf{\texttt{reg y x,
cluster(group) vce(jackknife)}} commands. However, many samples are
not well-behaved, in the sense that the regressor matrices for some of
the omit-one-cluster subsamples may not have full rank. We will refer
to such subsamples, rather informally, as ``singular subsamples.''

Whenever there are singular subsamples, \textbf{\texttt{summclust}}
calculates two standard errors. One of these drops the singular
subsamples, as the native Stata commands do. The other uses a
generalized inverse. \textbf{\texttt{summclust}} provides guidance as
to which standard error is likely to be more reliable. When
\textbf{\texttt{regtable}} is specified, and singular subsamples are
present, two versions of the regression table are displayed.
Similarly, if \textbf{\texttt{jackknife}} is specified and there are
singular subsamples, four different standard errors are shown, either
CV$_{\tn3}$ or CV$_{\tn3{\rm J}}$, combined with either the
generalized inverse or one that drops the singular subsamples.

\textbf{\texttt{nograph}} suppresses creation of the figure, which is
otherwise shown by default. The figure shows four scatter plots:
leverage against observations per cluster, partial leverage against
observations per cluster, leverage against omit-one-cluster
coefficients, and partial leverage against omit-one-cluster
coefficients. This figure can be quite informative, but it is
computationally costly to produce. We recommend invoking this option
after the figure has been inspected.

When \textbf{\texttt{jackknife}} is specified,
\textbf{\texttt{regtable}} uses the CV$_{\tn3{\rm J}}$ estimates to
produce the regression table. Otherwise, CV$_{\tn3}$ estimates are
used.




\subsection{Illustration with \texttt{nlswork}}

To illustrate \texttt{summclust}'s functionality and syntax, we
consider a simple example using the online dataset \texttt{nlswork},
which contains a sample of women who were 14--26 years of age in 1968
from the National Longitudinal Survey of Young Working Women.  For the
purposes of this exercise, we restrict the sample to individuals who
are 20 to 40 years old.

We estimate a simple Mincer regression using the \texttt{nlswork}
dataset to see whether there is a marriage premium for wages. The
variable \texttt{msp} is equal to 1 if the person is married and
cohabits with their spouse, and equal to 0 otherwise. For the purposes
of this example, we cluster by industry. The following code opens the
dataset and estimates the regression using \texttt{Stata}'s
\texttt{regress} command:

\begin{small}
\begin{verbatim}
webuse nlswork, clear
keep if inrange(age,20,40)
reg ln_wage i.grade i.age i.birth_yr union race msp, cluster(ind)
\end{verbatim}
\end{small}

\noindent The \texttt{Stata} output from the command above provides
CV$_{\tn1}$ standard errors. Alternatively, we can estimate
CV$_{\tn3}$ and CV$_{\tn3{\rm J}}$ standard errors using this code:

\begin{small}
\begin{verbatim}
reg ln_wage i.grade i.age i.birth_yr union race msp, cluster(ind) vce(jackknife, mse)
reg ln_wage i.grade i.age i.birth_yr union race msp, cluster(ind) vce(jackknife)
\end{verbatim}
\end{small}


\noindent When either of these commands is run, \texttt{Stata}
displays the warning ``Note: One or more parameters could not be
estimated in 2 jackknife replicates; standard-error estimates include
only complete replications.''
 
The coefficient on \texttt{msp} and two or three standard errors can
also be obtained using \texttt{summclust}. The basic command is:

\begin{small}
\begin{verbatim}
summclust ln_wage msp union race, fevar(grade age birth_yr) cluster(ind) 
\end{verbatim}
\end{small}

\noindent This code results in the default output from
\texttt{summclust}, which is mostly contained in two tables. The first
one includes the coefficient on the second variable in the varlist (in 
this case \texttt{msp}), the CV$_{\tn1}$ and CV$_{\tn3}$ standard 
errors for this coefficient, and the associated $t$-statistics, 
$P$~values, and confidence intervals. In this case, \texttt{summclust} 
also displays a warning about singular subsamples and thus produces two
``Regression Output'' tables. The standard errors in the table which 
drops singular subsamples match those produced natively in Stata.

\begin{small}
\begin{verbatim}
Cluster summary statistics for msp when clustered by ind_code.
There are 17395 observations within 12 ind_code clusters.
\end{verbatim}
\end{small}

\begin{small}
\begin{verbatim}
Regression Output
s.e. |      Coeff   Sd. Err.   t-stat  P value    CI-lower    CI-upper
------+---------------------------------------------------------------- 
 CV1 |  -0.026940   0.008248  -3.2663   0.0075   -0.045093   -0.008787
 CV3 |  -0.026940   0.011150  -2.4161   0.0342   -0.051481   -0.002399
-----------------------------------------------------------------------
\end{verbatim}
\end{small}

\begin{small}
\begin{verbatim}
Regression Output -- Dropping Singular Omit-One-Cluster Subsamples
s.e. |      Coeff   Sd. Err.   t-stat  P value    CI-lower    CI-upper
------+---------------------------------------------------------------- 
CV3  |  -0.026940   0.006701  -4.0200   0.0030   -0.042099   -0.011780
-----------------------------------------------------------------------
\end{verbatim}
\end{small}



\noindent In the first table for this example, the CV$_{\tn1}$ and
CV$_{\tn3}$ standard errors are noticeably different, with the latter
being considerably larger. However, in the second table, where the two
singular subsamples are dropped, the CV$_{\tn3}$ standard error
becomes much smaller.

The ``Cluster Variability'' table from \texttt{summclust} (below)
provides insight into what is happening. It reports summary statistics
for $N_g$, $L_g$, $L_{gj}$, and~$\hat\beta_{j}^{(g)}$. Whenever
singular subsamples are dropped, two sets of statistics are shown 
for~$\hat\beta_{j}^{(g)}$. The first (second-last column) uses all the
jackknife subsamples with a generalized inverse standard error. The
second (final column) uses only the non-singular subsamples. We can
see that the largest value of $\hat\beta_{j}^{(g)}$ is considerably
smaller (and therefore more different from the other values) when none
of the subsamples is dropped. This explains why the CV$_{\tn3}$
standard error is larger in the first table above than in the second
one.

\vfill
\eject

\begin{small}
\begin{verbatim}
Cluster Variability   
Statistic |       Ng    Leverage   Partial L.  all bet~g   kept be~g  
----------+---------------------------------------------------------
      min |    35.00    0.085945     0.000700  -0.032772   -0.032772  
       q1 |   144.50    0.633594     0.004399  -0.027655   -0.027917  
   median |   905.00    2.794231     0.038554  -0.026891   -0.027082  
     mean |  1449.58    4.583333     0.083333  -0.026398   -0.027571  
       q3 |  2112.50    6.190322     0.105043  -0.025268   -0.026587  
      max |  5736.00   17.008305     0.353148  -0.019198   -0.024202  
   -----------+-----------------------------------------------------
  coefvar |     1.19    1.166238     1.320154   0.131277    0.074100  
\end{verbatim}
\end{small}

\noindent It is evident from this table that the clusters are
extremely heterogeneous. The largest cluster contains almost
one\tkk-third of the sample and is 167 times the size of the smallest.
There are also extreme differences in both leverage and partial
leverage across clusters. The ratio of the largest to the smallest
value is 198 for leverage and 504.5 for partial leverage. The sum of
the leverages is $12 \times 4.583333 = 55$, which is the number of
estimated coefficients. Although both sets of $\hat\beta_{j}^{(g)}$
vary quite a bit, dropping one cluster never changes the sign of the
coefficient.

The option \texttt{fevar} is used when there are factor variables,
which would be specified as \texttt{i.varname} in conventional
\texttt{Stata} syntax. In the above example, the arguments to
\texttt{fevar} are \texttt{grade}, \texttt{age}, and
\texttt{birth\_yr}. For each argument, a set of temporary dummy
variables is created. These dummy variables are included in the
regression, and there is no constant term if they are present.

The sample code above does not illustrate several additional options.
The most important of these is the \texttt{absorb} option, which
operates like \texttt{fevar}. It treats its argument, a single
variable, as an additional factor variable to include in the set of
regressors. \texttt{absorb(varname)} can be used when including
\texttt{i.varname} in a regression would result in many fixed effects.
Speed can often be increased, perhaps substantially, by partialing out
the absorbed fixed effects from the dependent and all the independent
variables. It is advisable to use \texttt{absorb} rather than
\texttt{fevar} whenever their argument corresponds to a set of cluster
fixed effects, since the elements of $\hat\bbeta^{(g)}$ that
correspond to the fixed effects cannot be identified in that case; see
\Cref{sec:influential}.

The \texttt{absorb} option should be used with care. Partialing out
fixed effects is valid for the measures of leverage and influence and
for the jackknife variance matrices only when the absorbed variable
yields fixed effects that can be partialed out on a cluster-by-cluster
basis. That is, \texttt{absorb} should only be used for straight
cluster fixed effects or for fixed effects at a finer level, such as
state $\times$ year fixed effects for a panel with clustering at the
state level. It is not valid to partial out fixed effects that are not
limited to a single cluster. In that case, the $\hat\bbeta^{(g)}$ and
quantities based on them would be different for the original data and
the data after partialing out, because the partialed-out observations
for the \th{g} cluster would depend on other clusters as well.
Accordingly, \texttt{summclust} checks to ensure that the clustering
variable is invariant within each value of the absorbed variable. When
it is not invariant, a warning is displayed, and the values of  $L_{g}$,
$L_{gj}$, $\hat\beta_{j}^{(g)}$\tn, CV$_{\tn3}$, and CV$_{\tn3{\rm J}}$ are 
not available.






To see the difference between \texttt{fevar} and \texttt{absorb}, we
can estimate an expanded regression that includes industry fixed
effects. Consider the following two commands:

\begin{small}
\begin{verbatim}
summclust ln_wage msp union race, fevar(grade age birth_yr ind) cluster(ind) 
summclust ln_wage msp union race, fevar(grade age birth_yr) absorb(ind) cluster(ind) 
\end{verbatim}
\end{small}

\noindent For the command which uses \texttt{fevar} for all the
categorical variables, some of the output is

\begin{small}
\begin{verbatim}
Regression Output
  s.e. |      Coeff   Sd. Err.   t-stat  P value    CI-lower    CI-upper
-------+----------------------------------------------------------------
   CV1 |  -0.018955   0.007014  -2.7025   0.0206   -0.034392   -0.003517
   CV3 |  -0.018955   0.007586  -2.4987   0.0296   -0.035651   -0.002258
------------------------------------------------------------------------
\end{verbatim}
\end{small}

\noindent Because every one of the jackknife subsamples is singular,
only the results based on the generalized inverse are reported. In
contrast, when \texttt{absorb} is used for the industry fixed effects,
the corresponding output is instead

\begin{small}
\begin{verbatim}
Regression Output
  s.e. |      Coeff   Sd. Err.   t-stat  P value    CI-lower    CI-upper
-------+----------------------------------------------------------------
   CV1 |  -0.018955   0.007014  -2.7025   0.0206   -0.034392   -0.003517
   CV3 |  -0.018955   0.007586  -2.4987   0.0296   -0.035651   -0.002258
------------------------------------------------------------------------

Regression Output -- Dropping Singular Omit-One-Cluster Subsamples
------------------------------------------------------------------------
  s.e. |      Coeff   Sd. Err.   t-stat  P value    CI-lower    CI-upper
   CV3 |  -0.018955   0.004173  -4.5418   0.0014   -0.028396   -0.009514
------------------------------------------------------------------------
\end{verbatim}
\end{small}

\noindent These two tables highlight a key reason for using
\texttt{absorb}. Because only two of the jackknife subsamples are
singular, \texttt{summclust} is able to report both standard errors.
Observe that, when all 12 jackknife samples are used, the standard
errors are the same regardless of whether industry fixed effects are
specified using \texttt{fever} or \texttt{absorb}.

Whether we use \texttt{fevar} or \texttt{absorb} leads to somewhat
different output for the measures of cluster variability.



\begin{small}
\begin{verbatim}
Cluster Variability [using fevar]
 Statistic |       Ng      Leverage     Partial L.  beta no g    
-----------+-------------------------------------------------
       min |    35.00      1.079703       0.000276  -0.021394    
        q1 |   144.50      1.617131       0.003970  -0.020316    
    median |   905.00      3.752372       0.033630  -0.019050    
      mean |  1449.58      5.500000       0.083333  -0.018880    
        q3 |  2112.50      7.066207       0.092329  -0.018852    
       max |  5736.00     17.728424       0.382133  -0.012367    
-----------+-------------------------------------------------
   coefvar |     1.19      0.957329       1.422090   0.126464  	
\end{verbatim}
\end{small}

\begin{small}
\begin{verbatim}
Cluster Variability [using absorb]
 Statistic |       Ng    Leverage   Partial L.  all bet~g   kept be~g
-----------+---------------------------------------------------------
       min |    35.00    0.079703     0.000700  -0.021394   -0.021394  
        q1 |   144.50    0.617131     0.004399  -0.020316   -0.020601  
    median |   905.00    2.752372     0.038554  -0.019050   -0.019281  
      mean |  1449.58    4.500000     0.083333  -0.018880   -0.019538  
        q3 |  2112.50    6.066207     0.105044  -0.018852   -0.019028  
       max |  5736.00   16.728424     0.353143  -0.012367   -0.016767  
-----------+---------------------------------------------------------
   coefvar |     1.19    1.170068     1.320148   0.126464    0.061639  	
\end{verbatim}
\end{small}

\noindent The $\hat\beta_j^{(g)}$ when all clusters are retained are
identical for both options. But since there are two singular
subclusters, there are two versions of the $\hat\beta_j^{(g)}$ for the
\texttt{fevar} results. 




The leverage estimates are also smaller when we use the
\texttt{absorb} option. Recall that, for the original model with no
industry fixed effects, the leverages summed to~55. In the first case
just above, where the industry fixed effects are included as
regressors in \texttt{fevar}, the regression has 66 coefficients, and
the leverages therefore sum to $12 \times 5.5 = 66$. In the second
case, where the industry fixed effects are partialed out using
\texttt{absorb}, the regression has 54 coefficients, and the leverages
therefore sum to $12 \times 4.5 = 54$. Thus for the first case, each
of the leverages is larger than the corresponding one for the second
case by precisely~1.



\bigskip

\noindent\textbf{\texttt{\underline{Examples}}}


\noindent In the examples that follow, we include the \texttt{nograph}
option to reduce computational time.

\noindent This example illustrates the \texttt{jackknife} and
\texttt{table} options:

\begin{small}
\begin{verbatim}
summclust ln_wage msp union race, fevar(grade age birth_yr) cluster(ind) nog///
   jack table
\end{verbatim}
\end{small}

\begin{small}
\begin{verbatim}
Regression Output
  s.e. |      Coeff   Sd. Err.   t-stat  P value    CI-lower    CI-upper
-------+----------------------------------------------------------------
   CV1 |  -0.026940   0.008248  -3.2663   0.0075   -0.045093   -0.008787
   CV3 |  -0.026940   0.011150  -2.4161   0.0342   -0.051481   -0.002399
  CV3J |  -0.026940   0.011004  -2.4482   0.0324   -0.051160   -0.002720
------------------------------------------------------------------------
\end{verbatim}
\end{small}

\noindent In addition to the two standard tables, it displays
the following table:

\begin{small}
\begin{verbatim}
Cluster by Cluster Statistics
  ind_code |     Ng      Leverage     Partial L.  beta no g    
-----------+-------------------------------------------------
         1 |      119      0.581881       0.002825  -0.026959    
         2 |       35      0.085945       0.000700  -0.027206    
         3 |      170      0.685307       0.005341  -0.026823    
         4 |     3451     12.753229       0.241651  -0.021861    
         5 |      974      2.448713       0.114532  -0.024202    
         6 |     2626      7.815303       0.095555  -0.027393    
         7 |     1599      4.565341       0.048163  -0.026587    
         8 |      513      2.494440       0.018808  -0.029519    
         9 |      836      3.131195       0.028945  -0.032772    
        10 |      114      0.336320       0.003457  -0.027917    
        11 |     5736     17.008305       0.353148  -0.019198    
        12 |     1222      3.094021       0.086874  -0.026333   
-------------------------------------------------------------
\end{verbatim}
\end{small}

\noindent This table makes it easy to see whether the high leverage
clusters are also the largest clusters. That is clearly the case here.
After running the program, this table is stored as the \texttt{Mata}
matrix \texttt{scall}.

To obtain summary statistics on the four (or five) measures of cluster
variability, we can use the \texttt{addmeans} option: 

\begin{small}
\begin{verbatim}
summclust ln_wage msp union race, fevar(grade age birth_yr) cluster(ind) nog add
\end{verbatim}
\end{small}

\noindent This command produces the following table:

\begin{small}
\begin{verbatim}
Alternative Sample Means and Ratios to Arithmetic Mean
                |          Ng     Leverage  Partial L.  all bet~g   kept be~g  
----------------+------------------------------------------------------------
  Harmonic Mean |     206.576     0.608440    0.004988          .           .  
 Harmonic Ratio |       0.143     0.132751    0.059853          .           .  
 Geometric Mean |     623.091     2.042731    0.025557          .           .  
Geometric Ratio |       0.430     0.445687    0.306684          .           .  
 Quadratic Mean |    2193.268     6.870062    0.134308   0.026605    0.027654  
Quadratic Ratio |       1.513     1.498923    1.611699  -1.007868   -1.003015  
-----------------------------------------------------------------------------
\end{verbatim}
\end{small}

\noindent Once again, we see that there is extreme variability across
the clusters. This is particularly noticeable for the ratio of the
harmonic mean to the arithmetic mean, which is between 0.125 and 0.143
for the cluster size, leverage, and partial leverage measures. Recall
that these ratios would be close to one if the clusters were
relatively homogeneous. This table is stored in \texttt{Mata}'s memory
as \texttt{bonus}.

To obtain estimates of the effective number of clusters, we can use
either the \texttt{gstar} option or the \texttt{rho()} option. The
former displays $G_j^*(0)$ and $G_j^*(1)$. The latter requires a
specified value of $\rho$ and displays $G_j^*(0)$ and $G_j^*(1)$ along 
with $G_j^*(\rho)$. When there are fixed effects at the cluster or
subcluster level, only $G_j^*(0)$ is reported.

For the \texttt{nlswork} example, the first option
may be called as:

\begin{small}
\begin{verbatim}
summclust ln_wage msp union race, fevar(grade age birth_yr) cluster(ind) nog gstar 
\end{verbatim}
\end{small}

\noindent This yields:

\begin{small}
\begin{verbatim}
Effective Number of Clusters
----------------------------
G*(0)  =  5.495
G*(1)  =  1.376
----------------------------
\end{verbatim}
\end{small}

\noindent The second option, using $\rho = 0.5$ as an 
illustration, may be called as:

\begin{small}
\begin{verbatim}
summclust ln_wage msp union race, fevar(grade age birth_yr) cluster(ind) nog rho(0.5)
\end{verbatim}
\end{small}

\noindent This yields:


\begin{small}
\begin{verbatim}
Effective Number of Clusters
----------------------------
G*(0)  =  5.495
G*(.5) =  1.433
G*(1)  =  1.376
----------------------------
\end{verbatim}
\end{small}

\noindent In this example, it is clear that the effective number of
clusters is substantially less than the actual number of clusters.
This provides more evidence that inference using the CV$_{\tn1}$ 
standard error together with the $t(G-1)$ distribution is likely to be
unreliable. These three quantities can be accessed in \texttt{Mata}'s
memory as \texttt{gstarzero}, \texttt{gstarrho}, and
\texttt{gstarone}, respectively.

By using the \texttt{regtable} option, one can display a modified
version of the regression table, which is similar to the default
output from Stata's \texttt{regress} command. The command is:

\begin{small}
\begin{verbatim}
summclust ln_wage msp union race, fevar(grade age birth_yr) cluster(ind) nog regtable 
\end{verbatim}
\end{small}

\noindent When there are singular subsamples, two versions of this
table will be displayed. In this example, the table is quite long, so
we do not reproduce it here.

\subsection{List of Stored Results}
\label{subsec:stored}

All the results that are displayed as output can also be found in
\texttt{Mata}'s memory. To access one of these after running
\texttt{summclust}, simply add the following line:

\begin{small}
\begin{verbatim}
mata: object_name
\end{verbatim}
\end{small}

\noindent The \verb+object_name+ can take one of the following values:
\medskip

\hangpara {\tt cvstuff}: This matrix stores the table with the title 
``Regression Output''. It is $2\times6$ when the \texttt{jackknife} 
option is not used (the default), and $3\times6$ when 
\texttt{jackknife} is used.

\hangpara {\tt scall}: This matrix stores the $G\times4$ table created
by the \texttt{table} option with the title ``Cluster by Cluster
Statistics''.

\hangpara {\tt bonus}: This $6\times4$ matrix contains the alternative
sample means and their ratios to the arithmetic mean created by the
\texttt{addmeans} option.

\hangpara {\tt gstarzero}: This scalar contains $G^*(0)$ created by
the \texttt{gstar} or \texttt{rho} options.

\hangpara {\tt gstarone}: This scalar contains $G^*(1)$ created by
the \texttt{gstar} or \texttt{rho} options.

\hangpara {\tt gstarrho}: This scalar contains $G^*(\rho)$ created by
the \texttt{rho} option.

\hangpara {\tt regresstab}: This matrix contains the table shown when 
the regtable option is specified. 

\bigskip
\noindent Scalars within matrices can be referenced on a cell-by-cell 
basis. For example, the CV$_{\tn3}$ standard error is stored in the 
second row and second column of \texttt{cvstuff}, and to display it one 
can enter the following command:

\begin{small}
\begin{verbatim}
mata: cvstuff[2,2]
\end{verbatim}
\end{small}

\noindent Additionally, several results are available as scalars or 
matrices in return memory using \texttt{r()}. The available scalars
are: \medskip

\hangpara {\tt beta}: The estimate $\hat\beta$ for the coefficient of 
interest.

\hangpara {\tt cv1se}: The CV$_{\tn1}$ standard error for the 
coefficient of interest.

\hangpara {\tt cv1t}: The CV$_{\tn1}$ $t$-statistic for the coefficient
of interest.

\hangpara {\tt cv1p}: The $P$ value for the null hypothesis that 
$\beta = 0$ for the coefficient of interest using the CV$_{\tn1}$ 
standard error.

\hangpara {\tt cv1lci}: The lower bound of the 95\% confidence interval
for $\beta$ using the CV$_{\tn1}$ standard error.

\hangpara {\tt cv1uci}: The upper bound of the 95\% confidence  interval
for $\beta$ using the CV$_{\tn1}$ standard error.

\hangpara {\tt gstarzero}: The effective number of clusters for the 
coefficient of interest using $\rho=0$.

\hangpara {\tt gstarone}: The effective number of clusters for the 
coefficient of interest using $\rho=1$.

\hangpara {\tt gstarrho}: The effective number of clusters for the 
coefficient of interest using the value of $\rho$ specified in 
\texttt{rho($\rho$)}.

\bigskip \noindent The standard error, $t$-statistic, $P$~value, and
confidence interval bounds are also available for the CV$_{\tn3}$ and 
CV$_{\tn3{\rm J}}$ standard errors. To access these, replace ``1'' in
the above with either ``3'' or ``3J''; for example, the $P$~value 
using CV$_{\tn3{\rm J}}$ is available in \texttt{cv3Jp}.  In the event
of singular subsamples, there are two versions of the CV$_{\tn3}$ or
CV$_{\tn3{\rm J}}$ results. The ones where singular subsamples have
been dropped have a suffix of `drop'. For instance, \texttt{cv3sedrop}
is used instead of \texttt{cv3se}.


\bigskip
\noindent The available matrices are:  
\medskip

\hangpara {\tt ng}: This $G\times1$ matrix contains the number of 
observations, $N_g$, for each cluster.

\hangpara {\tt leverage}: This $G\times1$ matrix contains the leverage, 
$L_g$, for each cluster.

\hangpara {\tt partlev}: This $G\times1$ matrix contains the partial 
leverage, $L_{gj}$, for each cluster.

\hangpara {\tt betanog}: This $G\times1$ matrix contains the 
$\hat\beta_j^{(g)}$ for each cluster.

\section{Empirical Example}
\label{sec:empirical}




We consider an empirical example from \citet*{BG_2019}, which studies
an experiment where retail firms were randomly assigned to enter one of
72 different geographic markets (in Spanish, mercados), within the
Dominican Republic. After randomization, 21 markets had no entrants
and so were in the control group, 18 had one entrant, another 18 had
two, and the remaining 15 had three. The primary analysis only
distinguishes between the 51 treated markets and the 21 control
markets. The number of observations (stores) per market varies from 20
to~55.

This example is interesting because conventional wisdom
\citep*[e.g.,][]{MNW-guide} suggests that, with 72 clusters that do
not vary much in size, and with neither few treated nor few control
clusters, inference based on CV$_{\tn1}$ standard errors and the
$t(71)$ distribution should work well. However, our leverage measures
suggest otherwise, and alternative inference methods yield noticeably
different results.

The model we estimate is
\begin{equation}
Y_{sd} = \alpha + \gamma Z_d + \biX_{sd}\tk\bbeta + \epsilon_{sd}.
\label{mainreg}
\end{equation}
Here $s$ indexes stores, and $d$ indexes markets. The treatment
variable $Z_d$ equals~1 if market~$d$ is treated (there was entry)
and~0 if it was a control (there was no entry). The coefficient of
interest is $\gamma$, which measures the causal effect of increased
competition on an outcome $Y$\tn. We focus on just one of several
outcomes, namely, the log of demeaned prices after treatment. The
results from this regression are found in Table~5, Panel~B, column~4,
row~1 of \citet*{BG_2019}. The table states that there are 72 clusters
and $2,\tn025$ observations; however, the replication dataset that we
use contains just $1,\tn926$ observations.



Regression \eqref{mainreg} includes 17 control variables in the row
vector $\biX_{sd}$. These are the first lag of the outcome variable, 
the number of retailers in each district pre\tkk-treatment, a lagged
quality index, eight province fixed effects, total district
beneficiaries of a conditional cash transfer program, percent 
beneficiaries of that program, average income in the market, two 
market education measures, and a binary indicator for the urban 
status of the market. Thus the total number of regressors is~19.




\setcounter{table}{0}

\begin{table}[tp]
\caption{Estimates of the Treatment Effect}
\label{tab:ests}
\vspace*{-0.5em}
\begin{tabular*}{\textwidth}{@{\extracolsep{\fill}}lcccc}
\toprule
Method &$\hat\gamma$ &Standard Error &$P$ value &Confidence Interval \\
\midrule
CV$_{\tn1}$ &$-0.01469$ &$0.007243$ &$0.0461$ &[$-0.02913$,
$-0.00025$]\\
CV$_{\tn2}$ &$-0.01469$ &$0.008078$ &$0.0730$ &[$-0.03080$, 
\phantom{+}$0.00142$]\\
CV$_{\tn3}$ &$-0.01469$ &$0.009090$ &$0.1105$ &[$-0.03281$,
\phantom{+}$0.00343$]\\
CV$_{\tn3{\rm J}}$ &$-0.01469$ &$0.009087$ &$0.1104$ &[$-0.03281$,
\phantom{+}$0.00343$]\\
WCR-C bootstrap &$-0.01469$ & &$0.0891$ &[$-0.03121$,
\phantom{+}$0.00243$]\\
WCR-S bootstrap &$-0.01469$ & &$0.0913$ &[$-0.03121$,
\phantom{+}$0.00254$]\\
\bottomrule
\end{tabular*}
\vskip 6pt
{\footnotesize \textbf{Notes:} There are $N=1,\tn926$ observations and
$G=72$ clusters. The two WCR bootstraps use $B=999,\tn999$ and a seed
of $56,\tn829,\tn046$. WCR-C is the classic WCR bootstrap of
\citet*{CGM_2008}, and WCR-S is the ``score'' variant proposed in
\citet*{MNW-bootknife}. It involves transforming the restricted
empirical scores in a way based on the jackknife, but it still uses
CV$_{\tn1}$. The bootstrap results were obtained using Version 4.2.0
of \texttt{boottest}.}
\end{table}




The OLS estimate of $\gamma$, its CV$_{\tn1}$ standard error, the $P$
value for a test that $\gamma=0$, and a .95 confidence interval are
shown in the first row of \Cref{tab:ests}. Allowing for different
numbers of reported digits, these estimates accord with the ones in
\citet*{BG_2019}. The estimate of $-0.01469$ has the expected sign 
(average prices declined). However, the $P$ value is just slightly less 
than 0.05, and the confidence interval barely excludes~0.

\begin{table}[tp]
\caption{Leverage and Partial Leverage for $\hat\gamma$}
\label{tab:comp_sum}
\vspace*{-0.5em}
\begin{tabular*}{\textwidth}{@{\extracolsep{\fill}}lcccc}
\toprule
Statistic & $N_g$   & Leverage & Partial Leverage & $\hat\gamma^{(g)}$ \\
\midrule
Minimum   & 20\phantom{.0} & 0.130842 & 0.000099  & $-0.017550$ \\
First quartile & 24\phantom{.0} & 0.204104 & 0.003166  & $-0.015089$ \\
Median    & 26\phantom{.0} & 0.235813 & 0.009001  & $-0.014791$ \\
Mean      & \phantom{0}26.75 & 0.263889 & 0.013889  & $-0.014663$ \\
Third quartile & 27\phantom{.0} & 0.292042 & 0.020926  & $-0.014070$ \\
Maximum   & 55\phantom{.0} & 0.737797 & 0.064242  & $-0.010723$ \\
Coef.\ of variation &\phantom{00}0.21 &0.388686 &1.059813
  &$\phantom{-}0.074061$ \\
\bottomrule
\end{tabular*}
\vskip 6pt
{\footnotesize \textbf{Notes:} There are $N=1,\tn926$ observations and
$G=72$ clusters. The effective numbers of clusters are $G^*_{\gamma}(0) 
= 34.16$ and $G^*_{\gamma}(1) = 33.33$.}
\end{table}

We next use the \texttt{summclust} package to calculate the
cluster-level characteristics of the model and dataset. Some key ones
are reported in \Cref{tab:comp_sum}. It is evident that cluster sizes
are well balanced, varying from 20 to 55, with the first and third
quartiles equal to 24 and 27. However, both the leverages $L_g$ and
the partial leverages $L_{g1}$ vary considerably. The former range
from 0.1308 to 0.7378, and the latter from 0.0001 to 0.0642. The
coefficients of variation are 0.3887 and 1.0598, respectively. The
latter is moderately large, although not enormous. The two values of
$G^*$ are slightly smaller than $G/2$, which also suggests that the
sample is not well balanced.

Most of the $\hat\gamma^{(g)}$ do not vary much, and thus their
coefficient of variation is small. However, the most extreme values
are notable. The estimate of~$\gamma$, which is $-0.01469$, could be
as small as $-0.01755$ or as large as $-0.01072$ if just one out of 72
clusters were dropped. 




These results suggest that CV$_{\tn1}$, the default CRVE, may not be
particularly reliable in this case. We therefore consider five
alternative procedures. The second, third, and fourth rows of
\Cref{tab:ests} report the CV$_{\tn2}$, CV$_{\tn3}$, and 
CV$_{\tn3{\rm J}}$ standard errors, along with the $P$~values and 
confidence intervals associated with them. The CV$_{\tn2}$ $P$~value is
noticeable larger than the CV$_{\tn1}$ one and suggests that the
estimate is not significant at the .05 level. The CV$_{\tn3}$ and
CV$_{\tn3{\rm J}}$ rows are almost identical. At $0.1105$, the
CV$_{\tn3}$ $P$~value does not even allow us to reject the null at the
.10 level. The fifth and six rows of \Cref{tab:ests} report two WCR
bootstrap $P$~values and the associated .95 confidence intervals. At
0.0891 and 0.0913, these are a bit smaller than the jackknife ones,
but they clearly do not allow us to reject the null hypothesis at the
.05 level.

In view of the reasonably large number of clusters and the fact that
cluster sizes do not vary much, the large discrepancy between the
results for CV$_{\tn1}$ and the other procedures may seem surprising.
However, it is not all that surprising when we note how much the
leverages and, especially, the partial leverages vary.

\begin{figure}[tb]
\begin{center}
\caption{\label{fig:summclust}{Example summclust Figure}}
\vskip -8pt
\includegraphics[width=0.90\textwidth]{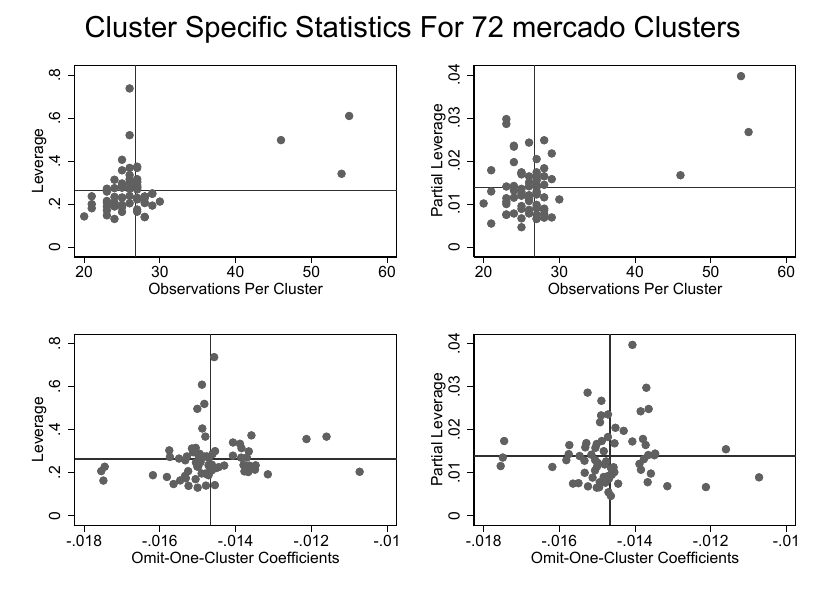}
\end{center}
\vskip -20pt
{\footnotesize \textbf{Notes:} A figure like this is always produced unless
the \texttt{nograph} option is specified. It plots both leverage and
partial leverage against cluster size and against the omit-one-cluster
coefficients for, in this case, 72 clusters specified by a variable
called ``mercado.''}
\end{figure}





By default, \texttt{summclust} produces a figure like Figure
\ref{fig:summclust}, with its title created by the program using the
name of the clustering variable, in this case ``mercado''. This figure
plots both leverage and partial leverage against the number of
observations per cluster and also against the omit-one-cluster
coefficients. These four subfigures may help to reveal the source of
cluster-level heterogeneity. For this example, neither the large
leverages nor the large partial leverages come exclusively from
clusters with large numbers of observations or extreme
omit-one-cluster coefficients.


\begin{figure}[tb]
\begin{center}
\caption{\label{fig:competition}{Partial Leverage vs Cluster Size}}
\vskip -8pt
\includegraphics[width=0.88\textwidth]{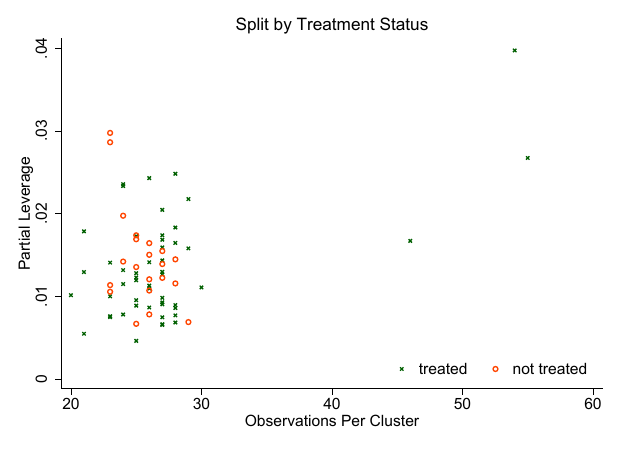}
\end{center}
\vskip -20pt
{\footnotesize \textbf{Notes:} The figure plots partial leverage against
cluster size for 72 clusters. A green X marks a treated cluster, and an
orange circle marks a control cluster.} 
\end{figure}

To explore what is driving the differences in partial leverage, we
create an additional scatter plot. \Cref{fig:competition} plots
partial leverage against the number of observations per cluster, with
different colors and symbols depending on whether or not a given
market (cluster) was treated. The figure has two interesting
features. The first is that the three rather large clusters have
fairly small partial leverage. The second is that the 12 clusters with
the highest partial leverage are all control markets. The first
result is quite surprising, since large clusters often tend to have
high leverage. But \Cref{fig:competition} makes it clear that there
is, in general, no simple relationship between cluster sizes and
partial leverage. The second result is not so surprising, because only
21 out of the 72 clusters are controls. Many of the control clusters
presumably have high partial leverage because control clusters are
relatively rare. See \eqref{eq:treatlev2} in \Cref{ex:treatconst} in 
the next section for an explanation.




\section{Simple Analytical Examples}
\label{sec:examples}

In this section, we discuss a number of simple examples in which it is
possible to calculate our measures of leverage and influence
analytically. These examples are quite revealing.

\begin{example}[Estimation of the mean]
\label{ex:mean}
Finding the sample mean is equivalent to performing a least-squares 
regression in which the only regressor is $x_i = 1$ for all 
$i=1,\ldots,N$\tn. In this case, it is easy to see that
$\biX_g^\top\!\biX_g = N_g$ and $\biX^\top\!\biX = N$\tn. Therefore,
\begin{equation}
L_g = \Tr(\biH_g) = \frac{N_g}{N} = \frac{N_g}{\sum_{h=1}^G N_h}.
\label{eq:conlev}
\end{equation}
In this simple case, cluster leverage is exactly proportional to
cluster size. In other cases, we can interpret leverage as a
generalization of cluster size that takes into account other types of
heterogeneity as well.

Evidently, $\hat\beta = \bar{y} = N^{-1}\sum_{g=1}^G N_g\tk\bar{y}_g$,
where $\bar{y}$ and $\bar{y}_g$ denote the sample average for the full
sample and for cluster $g$, respectively. This expression can be
rewritten as
\begin{equation}
\hat\beta = \sum_{g=1}^G\frac{N_g}{N}\tk\bar{y}_g =
\sum_{g=1}^G L_g\tk \hat\beta_g,
\label{eq:bhat}
\end{equation}
so that $\hat\beta$ is seen to be a weighted average of the $G$
estimates $\hat\beta_g=\bar y_g$, with the weight for each cluster
equal to its leverage. Similarly, we find that
\begin{equation}
\hat\beta^{(g)} = \frac{N}{N-N_g}\sum_{h \neq g} L_h\tk \hat\beta_h,
\label{eq:bhatg}
\end{equation}
where the first factor simply makes up for the fact that we are
summing over $G-1$ clusters instead of $G$ as in \eqref{eq:bhat}.
Subtracting \eqref{eq:bhat} from \eqref{eq:bhatg}, we conclude that
\begin{equation}
\hat\beta^{(g)} - \hat\beta 
= \frac{N_g}{N}\big(\hat\beta^{(g)} - \hat\beta_g\big)
= L_g\big(\hat\beta^{(g)} - \hat\beta_g\big).
\label{eq:bdiff}
\end{equation}
Therefore, cluster $g$ will be influential whenever omitting it yields
an estimate $\hat\beta^{(g)}$ that differs substantially from the
estimate $\hat\beta_g$ for cluster $g$ itself, especially when cluster
$g$ also has high leverage.
\end{example}

\begin{example}[Single regressor plus constant]
\label{ex:regressor}
Consider a regression design with a single regressor, $x_i$, and a 
constant term. Then
\begin{equation*}
\biX_g^\top\!\biX_g = \left[ {
\begin{array}{cc}
N_g & \sum_{i=1}^{N_g}x_{g,i} \\
\sum_{i=1}^{N_g}x_{g,i} & \sum_{i=1}^{N_g}x_{g,i}^2
\end{array} }
\right]
\quad \textrm{and} \quad
( \biX^\top\!\biX )^{-1} = 
\frac{1}{N^2 \hat\sigma_x^2}
\left[ {
\begin{array}{cc}
\sum_{i=1}^N x_i^2 & -\sum_{i=1}^N x_i \\
-\sum_{i=1}^N x_i & N
\end{array} }
\right]\!,
\end{equation*}
where $\hat\sigma_x^2$ denotes the sample variance of the $x_i$. After 
some algebra, we find that
\begin{equation}
L_g =
\frac{N_g}{N\hat\sigma_x^2}\big(\hat\sigma_x^2 + \hat\sigma_{x,g}^2
+ (\bar{x}_g-\bar{x})^2\big) ,
\label{eq:constlev}
\end{equation}
where $\bar{x}_g$ and $\hat\sigma_{x,g}^2$ denote the sample mean and 
sample variance of the $x_i$ within cluster~$g$. Expression
\eqref{eq:constlev} is a straightforward generalization of 
\eqref{eq:conlev}. The last two terms within the large parentheses are
the sample variance of the $x_{g,i}$ within cluster $g$ and the square
of the difference between $\bar{x}_g$ and $\bar{x}$. The sum of these
terms is the sample variance of the $x_{g,i}$ around $\bar{x}$ within
cluster~$g$. Thus cluster $g$ will have high leverage when the
variance of the $x_{g,i}$ around $\bar{x}$ within that cluster is
large relative to the variance $\hat\sigma_x^2$ for the full sample.
If everything except cluster sizes were perfectly balanced, $L_g$
would evidently reduce to $2N_g/N$\tn.

The partial leverage for $x$ is just
\begin{equation}
L_{g2} = \frac{N_g\big(\hat\sigma_{x,g}^2 +
(\bar{x}_g-\bar{x})^2\big)}{N\hat\sigma_x^2}\tk,
\label{eq:constplev}
\end{equation}
the total variation around $\bar{x}$ within cluster $g$ divided by the
total variation within the sample.  If everything except cluster sizes
were perfectly balanced, it would reduce to $N_g/N$\tn.
\end{example}

\begin{example}[Single regressor plus fixed effects]
\label{ex:regressorFE} Suppose there is a single regressor, $x_i$, and
there are cluster-level fixed effects, which have been partialed out.
In this case, we can write all quantities as deviations from their
cluster averages, and there is no distinction between leverage and
partial leverage. Then $\tilde\biX_g^\top\!\tilde\biX_g
=\sum_{i=1}^{N_g} (x_{g,i}-\bar{x}_g)^2 = N_g \hat\sigma_{x,g}^2$.
Similarly, $\tilde\biX^\top\!\tilde\biX = \sum_{g=1}^G N_g\tk
\hat\sigma_{x,g}^2$ is the average variance of the $x_i$ across all
clusters. We find that
\begin{equation}
L_g =
\frac{N_g\tk\hat\sigma_{x,g}^2}{\sum_{h=1}^G N_h\tk\hat\sigma_{x,h}^2},
\label{eq:felev}
\end{equation}
which is again a straightforward generalization of \eqref{eq:conlev}. 
The leverage of cluster $g$ is proportional to $N_g$ times the 
variance of the $x_{g,i}$ around $\bar{x}_g$. Thus, for example,
doubling the variance of the $x_{g,i}$ has the same effect on leverage
as doubling~$N_g$.

In this case, using \eqref{eq:felev}, it is easy to see that
\begin{equation}
\hat\beta = \frac{\sum_{g=1}^G N_g\tk \hat\sigma_{xy,g}}
{\sum_{g=1}^G N_g\tk \hat\sigma^2_{x,g}} =
\sum_{g=1}^G L_g \frac{\hat\sigma_{xy,g}}{\hat\sigma^2_{x,g}}
= \sum_{g=1}^G L_g \hat\beta_g,
\label{eq:bhatfe}
\end{equation}
where $\hat\sigma_{xy,g} = (1/N_g)\sum_{i=1}^{N_g} (x_{g,i} -
\bar{x}_g) (y_{g,i} - \bar y_g)$ is the sample covariance of $x_i$ and
$y_i$ within cluster~$g$. The rightmost expressions in \eqref{eq:bhat}
and \eqref{eq:bhatfe} are identical. In both cases, $\hat\beta$ is
seen to be a weighted average of the $G$ cluster estimates, with the
weight for each cluster equal to its leverage.

When cluster $g$ is omitted, we obtain
\begin{equation}
\hat\beta^{(g)} =
\frac{\sum_{h\ne g} N_h \hat\sigma_{xy,h}}
{\sum_{h\ne g} N_h \hat\sigma^2_{x,h}} =
\frac{\sum_{h\ne g} L_h \hat\beta_h}{\sum_{h\ne g}L_h},
\label{eq:bhatgfe}
\end{equation}
which would specialize to \eqref{eq:bhatg} if \eqref{eq:conlev} were
true. As before, $\hat\beta^{(g)}$ is a weighted average of the 
$\hat\beta_h$, with weights proportional to the $L_g$, which in this
case are also the partial leverages. Subtracting \eqref{eq:bhatfe}
from \eqref{eq:bhatgfe}, we find that
\begin{equation}
\hat\beta^{(g)} - \hat\beta = 
L_g\big(\hat\beta^{(g)} - \hat\beta_g\big),
\label{eq:betagdiff}
\end{equation}
which is formally identical to the rightmost expression in
\eqref{eq:bdiff}, although of course $L_g$ is defined in
\eqref{eq:felev} not \eqref{eq:conlev}. Cluster $g$ will be
influential whenever $\hat\beta^{(g)}$ differs substantially from the
estimate $\hat\beta_g$ for cluster $g$ itself, especially when cluster
$g$ also has high leverage.
\end{example}

\begin{example}[Treatment model with a constant term]
\label{ex:treatconst}
Now we specialize \Cref{ex:regressor} to the case in which the single
regressor is a treatment dummy denoted by $d_i$. Let $\bar{d}_g$ and
$\bar{d}$ denote the proportion of treated observations in cluster $g$
and in the sample, respectively. Then \eqref{eq:constlev} becomes
\begin{equation}
\label{eq:treatlev1}
L_g = \frac{N_g}{N} \Big(\frac{\bar{d}_g}{\bar{d}} +
\frac{1-\bar{d}_g}{1-\bar{d}}\Big).
\end{equation}
The first factor here is the relative size of the \th{g} cluster. The
second factor depends on how much $\bar{d}_g$ differs from $\bar{d}$.
When $\bar{d}_g = \bar{d}$, we see that $L_g = 2 N_g/N$. Otherwise,
the first term inside the parentheses causes leverage to be high
whenever $\bar{d}_g$ is large relative to $\bar{d}$, and the second
term causes leverage to be high whenever $\bar{d}_g$ is small relative
to~$\bar{d}$. As $\bar{d}$ increases for given $\bar{d}_g$, the first
term becomes smaller relative to the second term. Thus the \th{g}
cluster will tend to be influential either when it has a large
proportion of treated observations and the overall proportion is
small, or when it has a small proportion of treated observations and
the overall proportion is large.

We can also obtain the partial leverage of the treatment dummy for
this case. Expression \eqref{eq:constplev} simply becomes
\begin{equation}
\label{eq:treatplev1}
L_{g2} = \frac{N_g}{N} \Big(\frac{\bar{d}_g}{\bar{d}}
+ \frac{\bar{d} - \bar{d}_g}{1 - \bar{d}}\Big).
\end{equation}
Once again, the first factor is the relative size of the \th{g}
cluster. The second factor reduces to~1 when $\bar{d}_g=\bar{d}$, so
that $L_{g2} = N_g/N$ in that special case.

We can further specialize \eqref{eq:treatlev1} and
\eqref{eq:treatplev1} to models in which the treatment is applied at
the cluster level. Suppose that all observations in clusters $g = 1,
\ldots ,G_1$ are treated and no observations in the $G_0=G-G_1$
control clusters from $G_1+1$ to $G$ are treated. Then we find that
$\bar{d}_g = 1$ for $g = 1, \ldots ,G_1$, and $\bar{d}_g = 0$ for $g =
G_1 +1, \ldots ,G$. Inserting these into \eqref{eq:treatlev1} shows
that
\begin{equation}
\label{eq:treatlev2}
L_g = 
\begin{cases}
\frac{N_g}{N} \frac{1}{\bar{d}\addh8}
  & \textrm{for } g=1,\ldots,G_1, \\
\frac{N_g}{N} \frac{1}{1-\bar{d}\addh8\tk}
  & \textrm{for } g=G_1 +1,\ldots,G.
\end{cases}
\end{equation}
Inserting them into \eqref{eq:treatplev1} shows that
\begin{equation*}
L_{g2} =
\begin{cases}
\frac{N_g}{N} \frac{\bar{d}+1}{\bar{d}\addh8}
  & \textrm{for } g=1,\ldots ,G_1, \\
\frac{N_g}{N} \frac{\bar{d}}{1-\bar{d}\addh8\tk}
  & \textrm{for } g=G_1 +1,\ldots ,G.
\end{cases}
\end{equation*}
Thus any cluster tends to have high leverage if $N_g/N$ is large. A
treated cluster has high leverage and partial leverage if $\bar{d}$ is
small. Conversely, a control cluster has high leverage and partial
leverage if $\bar{d}$ is large.
\end{example}

\begin{example}[Treatment with fixed effects]
\label{ex:treatFE} Finally, we consider the case of cluster-level
fixed effects, where treatment is randomly applied at the individual
level. This is a special case of \Cref{ex:regressorFE}. We cannot
consider cluster fixed effects with cluster-level treatment, because
the treatment dummy would be invariant within clusters. We specialize
\eqref{eq:felev} and find that
\begin{equation}
L_g = \frac{N_g\tk \bar{d}_g (1-\bar{d}_g)}
{\sum_{h=1}^G N_h\tk \bar{d}_h (1-\bar{d}_h)} .
\label{eq:treatlev}
\end{equation}
Thus, as before, the leverage of cluster $g$, relative to the average
for the other clusters, is proportional to its size, $N_g$. It also
depends on the proportion of treated observations in the cluster. The
maximum (relative) leverage for cluster $g$ occurs at $\bar{d}_g=1/2$
and is symmetric around~$1/2$. The result \eqref{eq:betagdiff}
continues to hold. It tells us that cluster~$g$ will be influential
when its leverage \eqref{eq:treatlev} is large and $\hat\beta^{(g)}$
differs greatly from $\hat\beta_g$.
\end{example}

\section{Two\tkk-\tn Way Clustering}
\label{sec:twoway}

Up to this point, we have focused on one\tkk-way clustering. However,
it is also important to compute measures of leverage, partial
leverage, and influence when there is clustering in two or more
dimensions \citep*{CGM_2011}. In the simplest and most
commonly-encountered case, where there is two\tkk-way clustering, we
recommend computing the usual one\tkk-way measures of leverage,
partial leverage, and influence for each of the two clustering
dimensions. This requires calling \texttt{summclust} twice.


When the number of clusters in either dimension is small, or when the
data are seriously unbalanced in either dimension, conventional
inference based on a two\tkk-way version of CV$_{\tn1}$, together with
the $t(\min(G-1,H-1))$ distribution, can be seriously unreliable.
\citet*{MNW_2021} therefore suggests using the usual two\tkk-way
CV$_{\tn1}$ estimator and applying the original WCR bootstrap to the
dimension with the fewest clusters or the most unbalanced clusters.
Simulation evidence suggests that this often provides more reliable
inferences than the $t$ distribution, but these inferences may still
be problematic.


It may also be interesting to calculate measures of leverage, partial
leverage, and influence for the intersection of the two clustering
dimensions, especially when the number of non-empty intersections is
not large. This means calling \texttt{summclust} a third time. Suppose
there are two clustering dimensions, with $G$ clusters in the first
dimension and $H$ clusters in the second. Then the number of
intersection clusters is at most $GH$\tn, but it can be smaller if
some of the intersection clusters are empty. In order to use
\texttt{summclust} for the intersections, it is necessary to create a
new variable that uniquely identifies each of the non-empty
intersection clusters. Running \texttt{summclust} for this case may be
expensive when the number of non-empty intersections is large,
especially if $k$ is also large.

It is important to remember that, when \texttt{summclust} is invoked
three times for each of two clustering dimensions and their
intersection, the CV$_{\tn3}$ standard error that it reports for each
of the three cases is based on a different pattern of one\tkk-way
clustering. When two\tkk-way clustering is appropriate, none of these
standard errors is valid. However, what \texttt{summclust} reports can
be used to compute an asymptotically valid variance as
\begin{equation}
\widehat\var_{\rm 2W}(\hat\beta_j) =
\widehat\var_G(\hat\beta_j) + \widehat\var_H(\hat\beta_j) -
\widehat\var_{GH}(\hat\beta_j).
\label{eq:twoway}
\end{equation}
Here $\hat\beta_j$ is the OLS estimate of a coefficient of interest,
and the three estimated variances on the right-hand side of
\eqref{eq:twoway} are the squares of the CV$_{\tn3}$ or
CV$_{\tn3\rm{J}}$ standard errors reported by \texttt{summclust} for
clustering dimension $G$, clustering dimension $H$\tn, and the
intersection of the two clustering dimensions, respectively.

Asymptotically, the two\tkk-way variance $\var_{\rm 2W}(\hat\beta_j)$
should not be less than either of the one\tkk-way variances.
Therefore, if $\widehat\var_{\rm 2W}(\hat\beta_j)$ is less than either
$\widehat\var_G(\hat\beta_j)$ or $\widehat\var_H(\hat\beta_j)$, it
makes sense to replace it by the larger of those two variance
estimates. Doing this also eliminates the risk of having to take the
square root of a negative number.  The appropriate $t$ distribution
has $\min(G-1,H-1)$ degrees of freedom if $\widehat\var_{\rm
2W}(\hat\beta_j)$ is used and $G-1$ or $H-1$ degrees of freedom if it
is replaced by either $\widehat\var_G(\hat\beta_j)$ or
$\widehat\var_H(\hat\beta_j)$, respectively. We conjecture that,
especially when this is done, the two\tkk-way standard error based on
either jackknife estimator will yield more conservative, and generally
more reliable, inferences than the usual two\tkk-way standard error
based on CV$_{\tn1}$.




As we discuss in \Cref{sec:package}, it is often invalid to partial
out fixed effects when computing a jackknife CRVE. This can be
particularly tricky in the case of two\tkk-way clustering. For
example, suppose there are $G$ states and $H$ years. Then it may be
desirable to partial out the state fixed effects when computing
$\widehat\var_G(\hat\beta_j)$ but invalid to partial out the year
fixed effects. Similarly, it may be desirable to partial out the year
fixed effects when computing $\widehat\var_H(\hat\beta_j)$ but invalid
to partial out the state fixed effects. Finally, it is invalid to
partial out either set of fixed effects when computing
$\widehat\var_{GH}(\hat\beta_j)$. The \texttt{absorb} option of
\texttt{summclust} normally detects cases where partialing out is
invalid and refuses to display jackknife standard errors and several
other quantities.

\section{Simulation Experiments}
\label{sec:sims}


One of the reasons for calculating leverages and partial leverages is
to identify cases in which inference may be problematical. The
objective of the simulation experiments in this section is to see
whether the rejection frequencies for cluster-robust $t$-tests can be
predicted from the features of the $\biX$ matrix reported by
\texttt{summclust}. There are 3000 cases, each corresponding to a
particular $\biX$ matrix. For each case, we generate 10,000 values of
$\biy$ and use them to estimate rejection frequencies for $t$-tests or
bootstrap tests at the .05 level.

In the experiments, there are either 20 clusters and 2000 observations
or 30 clusters and 3000 observations. The cluster sizes $N_g$ are
determined by a parameter $\gamma\ge0$, as follows:
\begin{equation*}
N_g = \left[N\frac{\exp(\gamma g/G)}{\sum_{j=1}^G
\exp(\gamma j/G)}\right]\!,\;\; g=1,\ldots,G-1,
\end{equation*}
where $[\cdot]$ denotes the integer part of its argument, and $N_G = N
- \sum_{j=1}^{G-1} N_g$. As $\gamma$ increases, the cluster sizes
become increasingly unbalanced. The value of $\gamma$ is chosen
randomly from the $\U[2,4]$ distribution, so that the cluster sizes
tend to vary quite a lot. When $G=20$, the smallest cluster has
between 8 and 32 observations, and the largest has between 229 and
378. When $G=30$, the smallest cluster has between 7 and 32
observations, and the largest has between 237 and 396.

There are five regressors, one of which is the test regressor, plus a
constant term. The regressors equal either 0 or 1. With probability
$1-p_c$, all the observations in a cluster are~0. With probability
$p_c$, they randomly equal either 0 or 1, both with probability 0.5.
Thus, when $p_c=1$, all variation is at the individual level, and
leverage tends to be proportional to cluster sizes. As $p_c$ declines,
the samples become more unbalanced. In the experiments, the values of
$p_c$ are chosen to be 0.25, 0.30, 0.35, 0.40, 0.50, and 0.60, each
for one\tkk-sixth of the cases. Smaller values of $p_c$ tend to be
associated with larger discrepancies between actual rejection
frequencies and .05, the nominal level of the tests.

For each experiment, we obtain $12,\tn000$ estimated rejection
frequencies. One\tkk-quarter of these are based on CV$_{\tn1}$ and the
$t(G-1)$ distribution, one\tkk-quarter on CV$_{\tn3}$ and the $t(G-1)$
distribution, and one\tkk-quarter on each of the WCR-C and WCR-S
bootstraps. To predict these rejection frequencies, we use a
generalized additive model based on smoothing splines; see
\citet*[Section~7.7]{JWHT}. The base model can be written as
\begin{equation}
r_i = \beta_0 + f_1(V_{si}) + f_2(V^{1/2}_{si}) + \beta_1 G^*_{i0} +
u_i,
\label{eq:ssmodel}
\end{equation}
where $r_i$ is the rejection frequency for case~$i$. Here $V_{si}$
denotes $V_s(L_{{\bullet}j})$, the scaled variance of the partial
leverages $L_{gj}$ for the test regressor for case~$i$, $G^*_{i0}$
denotes $G^*_j(0)$ for the test regressor for case~$i$ (recall from
\Cref{sec:report} that it is a monotonically decreasing function of
the $L_{gj}$), and $f_1(\cdot)$ and $f_2(\cdot)$ are smoothing splines
with five degrees of freedom. Since everything on the right-hand side
of \eqref{eq:ssmodel} is a function of $V_{si}$, this model is simply
using the $V_{si}$ to predict the $r_i$ in a potentially nonlinear
way.

\begin{figure}[tb]
\begin{center}
\caption{Predicted rejection frequencies for asymptotic and bootstrap
tests at .05 level}
\label{fig:rejfreq}
\includegraphics[width=0.99\textwidth]{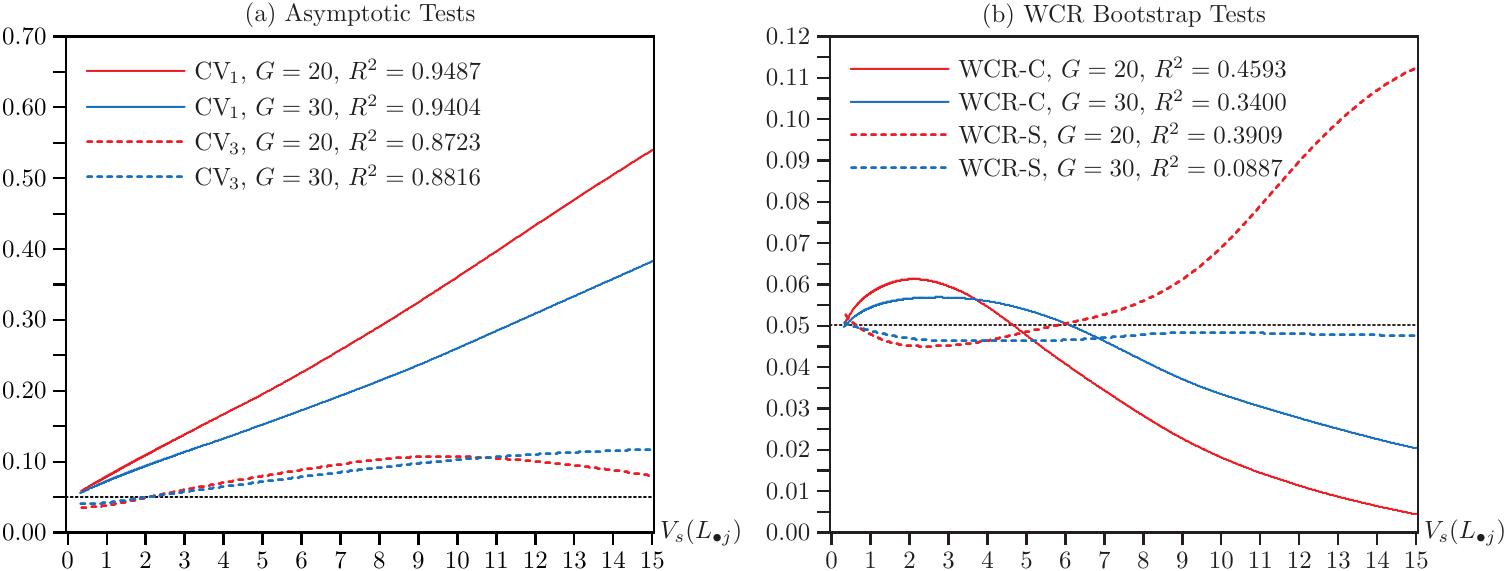}
\end{center}
{\footnotesize \textbf{Notes:} Each of the curves shows fitted values
from the generalized additive model \eqref{eq:ssmodel} that predicts
observed rejection frequencies, based on 10,000 replications, using
nonlinear functions of the $V_s(L_{{\bullet}j})$; see the text for
details. Bootstrap rejection frequencies are based on $B=399$.
WCR-C is the classic restricted wild cluster bootstrap, and WCR-S is
the score variant proposed in \citet*{MNW-bootknife}.}
\end{figure}

\Cref{fig:rejfreq} shows the fitted values from \eqref{eq:ssmodel},
which are predicted rejection frequencies, plotted against the scaled
variance of the partial leverages $L_{gj}$ for four methods of
inference and two sample sizes. Panel~(a) shows them for $t$-tests
based on both CV$_{\tn1}$ (solid lines) and CV$_{\tn3}$ (dashed lines)
for $G=20$ and $G=30$, and Panel~(b) shows them for WCR-C and
\mbox{WCR-S} bootstrap tests for the same two cases. The model seems
to fit quite well, at least for the asymptotic tests, as can be seen
from the values of $R^2$ reported for each of the curves. It also fits 
well for the bootstrap tests, and in fact it has smaller residuals
for them than for the asymptotic tests. The lower $R^2$ values for the
bootstrap tests simply reflect the fact that there is much less
variation to explain.


We can see from \Cref{fig:rejfreq} that $t$-tests based on CV$_{\tn1}$
often over-reject to an extreme degree. For the very smallest values
of $V_s(L_{{\bullet}j})$, the tests tend to over-reject modestly, with
predicted rejection frequencies of 0.058 for $G=20$ and 0.055 for
$G=30$. However, these then rise quite rapidly and almost linearly.
For $G=30$, there are four cases (out of 3000) for which
$V_s(L_{{\bullet}j})>15$. These are not shown in the figure, but the
approximately linear relationship continues to hold, and the fit for
these extreme cases is reasonably good.

In contrast, the $t$-tests based on CV$_{\tn3}$ tend to under-reject
for small values of $V_s(L_{{\bullet}j})$. For the very smallest
values, the predicted rejection frequencies are 0.033 for $G=20$ and
0.039 for $G=30$. Although it is not obvious from the figure, the
CV$_{\tn3}$ tests are predicted to under-reject somewhat more than
half the time, because, in our experiments, most values of
$V_s(L_{{\bullet}j})$ are quite small. As $V_s(L_{{\bullet}j})$
increases, rejection frequencies increase, although for $G=20$ they
start to decline again once $V_s(L_{{\bullet}j})$ exceeds about 9.6.
The predicted rejection frequencies never exceed 0.105 for $G=20$ and
0.118 for $G=30$. In a few cases (74 for $G=20$ and 5 for $G=30$), the
matrix that is inverted in \eqref{eq:delone} was singular for at least
one omit-one-cluster subsample. This happened whenever one of the
regressors took the same value for all observations in $G-1$ of the
clusters. These cases were dropped.


Panel~(b) of \Cref{fig:rejfreq} shows the fitted values from
\eqref{eq:ssmodel} for WCR-C and WCR-S bootstrap $t$-tests plotted
against the scaled variance of the~$L_{gj}$. Notice that the scale of
the vertical axis differs greatly from the one in Panel~(a). All
tests, especially the WCR-S ones, perform quite well for smaller
values of $V_s(L_{{\bullet}j})$. Except for WCR-S with $G=30$,
however, the rejection-frequency curves are not even close to being
linear. This is also the only case for which the fitted values do not
deviate greatly from 0.05 for large values of $V_s(L_{{\bullet}j})$.
In every other case, a large value of $V_s(L_{{\bullet}j})$ tends to
be associated with substantial levels of over-rejection or
under-rejection.

It is natural to ask whether we can improve the fit of
\eqref{eq:ssmodel} by adding additional explanatory variables that are
not simply functions of the $V_s(L_{{\bullet}j})$. The answer is that
we can. In particular, the variables $\bar a_{\rm
geo}(L_{{\bullet}j})$ and $G^*_j(1)$ are often significant when they
are added. However, the spline $f_1(V_{si})$ always remains highly
significant, even when many other regressors are included. Thus, at
least in these experiments, the scaled variance of the partial
leverages, which is the square of their coefficient of variation,
seems to be particularly revealing.

Based on these results, which are of course extremely dependent on the
way in which the regressors are generated, it seems sensible for
investigators to look at a number of different summary measures for
both leverage and partial leverage. That is why \texttt{summclust}
reports several of them. In this case, the most informative summary
measure appears to be the scaled variance, defined in
\eqref{eq:coefvar}, of the partial leverage measures $L_{gj}$, defined
in \eqref{eq:partial}, for the regressor of interest.
\texttt{summclust} reports the square root of this in the ``Coefvar''
line of the ``Cluster Variability'' table. In general, cluster-robust
inference seems to be most reliable when the partial leverages do not
vary greatly across clusters.

\section{Conclusions}
\label{sec:conc}


We have discussed a new \texttt{Stata} package called
\texttt{summclust} that is designed to summarize the cluster structure
of the dataset for a linear regression model with clustered
disturbances. Since the key unit of observation is the cluster, it
makes sense to examine measures of influence, leverage, and partial
leverage at the cluster level. These are easy to compute and are
conceptually very similar to the corresponding classic measures at the
observation level \citep*{BKW_1980,CH_1986}. The \texttt{summclust}
package calculates all of them and also reports a number of summary
statistics.

Our measure of influence at the cluster level can provide valuable
information about how empirical results depend on the data in the
various clusters. Investigators should be wary if dropping one or two
clusters changes the results dramatically. However, apart from such
cases, the most interesting quantities that \texttt{summclust}
calculates generally seem to be the partial leverages and measures
that summarize their distribution.

It has long been known that cluster-robust inference can be unreliable
when the number of clusters is small. More recent work, including
\citet*{MW-JAE,MW-EJ} and \citet*{DMN_2019}, has shown that it can
also be severely unreliable when cluster sizes vary a lot or when few
clusters are treated in the context of difference-in-differences and
other treatment models. In both of these cases, leverage and partial
leverage tend to vary greatly across clusters. It therefore seems
natural to use our measures of leverage and partial leverage as
diagnostic tools to identify datasets and regression designs in which
cluster-robust inference is likely to be challenging. Simulation
results in \Cref{sec:sims} suggest that the extent to which partial
leverage varies across clusters can be particularly informative. We
believe that investigators should always look at the summary
statistics reported by \texttt{summclust} and exercise caution
whenever they indicate substantial variation across clusters.

As we discuss in \Cref{sec:jack}, the computations needed for leverage
and influence are very similar to the ones needed to compute cluster
jackknife variance matrix estimators. The \texttt{summclust} package
therefore computes two very similar jackknife estimators, which we
refer to as CV$_{\tn3}$ and CV$_{\tn3{\rm J}}$, almost as a byproduct
of other computations. These are the same estimators that
\texttt{Stata} can produce using the \texttt{vce(jackknife,mse)} and
\texttt{vce(jackknife)} options. However, because \texttt{summclust}
is designed explicitly for linear regression models estimated by OLS, 
it is faster than using these \texttt{vce} options. Moreover, when
\texttt{summclust} is already being used to obtain cluster-level
measures of influence and leverage for diagnostic purposes, the
additional cost of computing the jackknife variance estimators is
minimal.

When the number of clusters is reasonably large and the variation of
leverage and partial leverage across clusters is small, we would
expect conventional inference based on CV$_{\tn1}$ standard errors to
perform well. If so, the CV$_{\tn3}$ standard errors reported by
\texttt{summclust} should be very similar to the CV$_{\tn1}$ standard
errors reported by one of \texttt{Stata}'s regression commands. When
this is the case, there is probably no need for investigators to worry
further about the reliability of their inferences. In many cases,
however, the CV$_{\tn3}$ and CV$_{\tn1}$ standard errors will differ
noticeably. This happens for the empirical example in
\Cref{sec:empirical}, where there are 72 clusters but partial leverage
varies a lot. In such cases, the CV$_{\tn3}$ standard errors are
almost certain to be more conservative, and very likely to be more
reliable, than the CV$_{\tn1}$ ones.





$P$~values and confidence intervals that are even more reliable can
often be obtained by using the restricted wild cluster bootstrap,
which is implemented natively with \texttt{wildbootstrap} in 
\texttt{Stata} 18 and in the package \texttt{boottest} \citep*{RMNW}.
Recent versions of that package implement the WCR-S bootstrap 
\citep*{MNW-bootknife} in addition to the classic WCR-C bootstrap. 
We strongly recommend that both variants be calculated whenever the
CV$_{\tn3}$ and CV$_{\tn1}$ standard errors disagree. When the two
bootstrap $P$~values agree, as they do for the empirical example in
\Cref{sec:empirical}, then it is probably safe to rely on either of 
them. When they disagree, then neither of them may be 
entirely reliable, but we would be inclined to use the one given by 
the WCR-S bootstrap.

Up to this point, everything in this section has been based on the
assumption that there is one\tkk-way clustering with a known
clustering structure. When more than one level of clustering is
plausible, investigators need to choose among them, and this can be
challenging; see the discussions in \citet*{MNW-guide,MNW-testing}.
The measures of leverage and influence produced by \texttt{summclust}
may be helpful in deciding at what level to cluster.


The current version of \texttt{summclust} is not explicitly designed
to handle two\tkk-way clustering. However, as we discuss in
\Cref{sec:twoway}, it can be called for each clustering dimension so
as to produce two sets of diagnostic statistics. If it is called three
times, once for each dimension and once for their intersection, then
it can also be used to compute two\tkk-way cluster jackknife variance
matrix estimators. At present, however, little is known about the
properties of these estimators.

\section*{Software Installation}
\label{sec:install}

To install the software files as they exist at the time of publication
of this article, type

\begin{verbatim}
   . net sj 23-4
   . net install st00!!  (to install program files, if available)
   . net get st00!!      (to install ancillary files, if available)
\end{verbatim}

The command \texttt{summclust} can be installed from the Statistical
Software Components archive by typing

\begin{verbatim}
   . ssc install summclust
\end{verbatim}
or from GitHub by typing
\begin{verbatim}
   net install summclust, ///
   from("https://raw.githubusercontent.com/mattdwebb/summclust/main/")
\end{verbatim}

\setlength{\bibsep}{1pt}
\bibliography{mnw-influence}

\begin{thebibliography}{}

\bibitem[\protect\citeauthoryear{Bell and McCaffrey}{Bell and
  McCaffrey}{2002}]{BM_2002}
Bell, R.~M. and D.~F. McCaffrey (2002).
\newblock Bias reduction in standard errors for linear regression with
  multi-stage samples.
\newblock {\em Survey Methodology\/}~{\em 28}, 169--181.

\bibitem[\protect\citeauthoryear{Belsley, Kuh, and Welsch}{Belsley
  et~al.}{1980}]{BKW_1980}
Belsley, D.~A., E.~Kuh, and R.~E. Welsch (1980).
\newblock {\em Regression Diagnostics}.
\newblock New York: Wiley.

\bibitem[\protect\citeauthoryear{Bester, Conley, and Hansen}{Bester
  et~al.}{2011}]{BCH_2011}
Bester, C.~A., T.~G. Conley, and C.~B. Hansen (2011).
\newblock Inference with dependent data using cluster covariance estimators.
\newblock {\em Journal of Econometrics\/}~{\em 165}, 137--151.

\bibitem[\protect\citeauthoryear{Broderick, Giordano, and Meager}{Broderick
  et~al.}{2021}]{BGM_2021}
Broderick, T., R.~Giordano, and R.~Meager (2021).
\newblock An automatic finite-sample robustness metric: {C}an dropping a little
  data change conclusions?
\newblock Ar{X}iv e-prints 2011.14999, London School of Economics.

\bibitem[\protect\citeauthoryear{Busso and Galiani}{Busso and
  Galiani}{2019}]{BG_2019}
Busso, M. and S.~Galiani (2019).
\newblock The causal effect of competition on prices and quality: {E}vidence
  from a field experiment.
\newblock {\em American Economic Journal: Applied Economics\/}~{\em 11},
  33--56.

\bibitem[\protect\citeauthoryear{Cameron, Gelbach, and Miller}{Cameron
  et~al.}{2008}]{CGM_2008}
Cameron, A.~C., J.~B. Gelbach, and D.~L. Miller (2008).
\newblock Bootstrap-based improvements for inference with clustered errors.
\newblock {\em Review of Economics and Statistics\/}~{\em 90}, 414--427.

\bibitem[\protect\citeauthoryear{Cameron, Gelbach, and Miller}{Cameron
  et~al.}{2011}]{CGM_2011}
Cameron, A.~C., J.~B. Gelbach, and D.~L. Miller (2011).
\newblock Robust inference with multiway clustering.
\newblock {\em Journal of Business \& Economic Statistics\/}~{\em 29},
  238--249.

\bibitem[\protect\citeauthoryear{Cameron and Miller}{Cameron and
  Miller}{2015}]{CM_2015}
Cameron, A.~C. and D.~L. Miller (2015).
\newblock A practitioner's guide to cluster-robust inference.
\newblock {\em Journal of Human Resources\/}~{\em 50}, 317--372.

\bibitem[\protect\citeauthoryear{Carter, Schnepel, and Steigerwald}{Carter
  et~al.}{2017}]{CSS_2017}
Carter, A.~V., K.~T. Schnepel, and D.~G. Steigerwald (2017).
\newblock Asymptotic behavior of a $t$ test robust to cluster heterogeneity.
\newblock {\em Review of Economics and Statistics\/}~{\em 99}, 698--709.

\bibitem[\protect\citeauthoryear{Chatterjee and Hadi}{Chatterjee and
  Hadi}{1986}]{CH_1986}
Chatterjee, S. and A.~S. Hadi (1986).
\newblock Influential observations, high-leverage points, and outliers in
  linear regression.
\newblock {\em Statistical Science\/}~{\em 1}, 379--416.

\bibitem[\protect\citeauthoryear{Chesher}{Chesher}{1989}]{Chesher_1989}
Chesher, A. (1989).
\newblock H\' ajek inequalities, measures of leverage and the size of
  heteroskedasticity robust tests.
\newblock {\em Econometrica\/}~{\em 57}, 971--977.

\bibitem[\protect\citeauthoryear{Conley, {Gon\c calves}, and Hansen}{Conley
  et~al.}{2018}]{CGH_2018}
Conley, T.~G., S.~{Gon\c calves}, and C.~B. Hansen (2018).
\newblock Inference with dependent data in accounting and finance applications.
\newblock {\em Journal of Accounting Research\/}~{\em 56}, 1139--1203.

\bibitem[\protect\citeauthoryear{Cook and Weisberg}{Cook and
  Weisberg}{1980}]{CW_1980}
Cook, R.~D. and S.~Weisberg (1980).
\newblock Characterizations of an empirical influence function for detecting
  influential cases in regression.
\newblock {\em Technometrics\/}~{\em 22}, 495--508.

\bibitem[\protect\citeauthoryear{Davidson and Mac\-Kinnon}{Davidson and
  Mac\-Kinnon}{1993}]{DM_1993}
Davidson, R. and J.~G. Mac\-Kinnon (1993).
\newblock {\em Estimation and Inference in Econometrics}.
\newblock New York: Oxford University Press.

\bibitem[\protect\citeauthoryear{Djogbenou, Mac\-Kinnon, and Nielsen}{Djogbenou
  et~al.}{2019}]{DMN_2019}
Djogbenou, A.~A., J.~G. Mac\-Kinnon, and M.~{\O}. Nielsen (2019).
\newblock Asymptotic theory and wild bootstrap inference with clustered errors.
\newblock {\em Journal of Econometrics\/}~{\em 212}, 393--412.

\bibitem[\protect\citeauthoryear{Efron}{Efron}{1979}]{Efron_79}
Efron, B. (1979).
\newblock Bootstrapping methods: Another look at the jackknife.
\newblock {\em Annals of Statistics\/}~{\em 7}, 1--26.

\bibitem[\protect\citeauthoryear{Hansen}{Hansen}{2022}]{Hansen-jack}
Hansen, B.~E. (2022).
\newblock Jackknife standard errors for clustered regression.
\newblock Working paper, University of Wisconsin.

\bibitem[\protect\citeauthoryear{Hansen and Lee}{Hansen and
  Lee}{2019}]{HansenLee_2019}
Hansen, B.~E. and S.~Lee (2019).
\newblock Asymptotic theory for clustered samples.
\newblock {\em Journal of Econometrics\/}~{\em 210}, 268--290.

\bibitem[\protect\citeauthoryear{Imbens and {Koles\'ar}}{Imbens and
  {Koles\'ar}}{2016}]{Imbens_2016}
Imbens, G.~W. and M.~{Koles\'ar} (2016).
\newblock Robust standard errors in small samples: Some practical advice.
\newblock {\em Review of Economics and Statistics\/}~{\em 98}, 701--712.

\bibitem[\protect\citeauthoryear{James, Witten, Hastie, and Tibshirani}{James
  et~al.}{2021}]{JWHT}
James, G.~M., D.~M. Witten, T.~J. Hastie, and R.~J. Tibshirani (2021).
\newblock {\em An Introduction to Statistical Learning\/} (Second ed.).
\newblock New York: Springer.

\bibitem[\protect\citeauthoryear{Lee and Steigerwald}{Lee and
  Steigerwald}{2018}]{LS_2018}
Lee, C.~H. and D.~G. Steigerwald (2018).
\newblock Inference for clustered data.
\newblock {\em Stata Journal\/}~{\em 18}, 447--460.

\bibitem[\protect\citeauthoryear{Mac\-Kinnon, Nielsen, and Webb}{Mac\-Kinnon
  et~al.}{2021}]{MNW_2021}
Mac\-Kinnon, J.~G., M.~{\O}. Nielsen, and M.~D. Webb (2021).
\newblock {\FO{1}}{W}ild bootstrap and asymptotic inference with multiway
  clustering.
\newblock {\em Journal of Business \& Economic Statistics\/}~{\em 39},
  505--519.

\bibitem[\protect\citeauthoryear{Mac\-Kinnon, Nielsen, and Webb}{Mac\-Kinnon
  et~al.}{2023a}]{MNW-guide}
Mac\-Kinnon, J.~G., M.~{\O}. Nielsen, and M.~D. Webb (2023a).
\newblock Cluster-robust inference: {A} guide to empirical practice.
\newblock {\em Journal of Econometrics\/}~{\em 232}, 272--299.

\bibitem[\protect\citeauthoryear{Mac\-Kinnon, Nielsen, and Webb}{Mac\-Kinnon
  et~al.}{2023b}]{MNW-bootknife}
Mac\-Kinnon, J.~G., M.~{\O}. Nielsen, and M.~D. Webb (2023b).
\newblock Fast jackknife and bootstrap methods for cluster-robust inference.
\newblock {\em Journal of Applied Econometrics\/}~{\em 38}, 671--694.

\bibitem[\protect\citeauthoryear{Mac\-Kinnon, Nielsen, and Webb}{Mac\-Kinnon
  et~al.}{2023c}]{MNW-testing}
Mac\-Kinnon, J.~G., M.~{\O}. Nielsen, and M.~D. Webb (2023c).
\newblock Testing for the appropriate level of clustering in linear regression
  models.
\newblock {\em Journal of Econometrics\/}~{\em 235}, 2027--2056.

\bibitem[\protect\citeauthoryear{Mac\-Kinnon and Webb}{Mac\-Kinnon and
  Webb}{2017a}]{MW-JAE}
Mac\-Kinnon, J.~G. and M.~D. Webb (2017a).
\newblock {\FO{1}}{W}ild bootstrap inference for wildly different cluster
  sizes.
\newblock {\em Journal of Applied Econometrics\/}~{\em 32}, 233--254.

\bibitem[\protect\citeauthoryear{Mac\-Kinnon and Webb}{Mac\-Kinnon and
  Webb}{2017b}]{MW-TPM}
Mac\-Kinnon, J.~G. and M.~D. Webb (2017b).
\newblock {\FO{2}}{P}itfalls when estimating treatment effects using clustered
  data.
\newblock {\em The Political Methodologist\/}~{\em 24}, 20--31.

\bibitem[\protect\citeauthoryear{Mac\-Kinnon and Webb}{Mac\-Kinnon and
  Webb}{2018}]{MW-EJ}
Mac\-Kinnon, J.~G. and M.~D. Webb (2018).
\newblock The wild bootstrap for few (treated) clusters.
\newblock {\em Econometrics Journal\/}~{\em 21}, 114--135.

\bibitem[\protect\citeauthoryear{Mac\-Kinnon and Webb}{Mac\-Kinnon and
  Webb}{2020}]{MW-RI}
Mac\-Kinnon, J.~G. and M.~D. Webb (2020).
\newblock Randomization inference for difference-in-differences with few
  treated clusters.
\newblock {\em Journal of Econometrics\/}~{\em 218}, 435--450.

\bibitem[\protect\citeauthoryear{Mac\-Kinnon and White}{Mac\-Kinnon and
  White}{1985}]{MW_1985}
Mac\-Kinnon, J.~G. and H.~White (1985).
\newblock Some heteroskedasticity consistent covariance matrix estimators with
  improved finite sample properties.
\newblock {\em Journal of Econometrics\/}~{\em 29}, 305--325.

\bibitem[\protect\citeauthoryear{Niccodemi, Alessie, Angelini, Mierau, and
  Wansbeek}{Niccodemi et~al.}{2020}]{NAAMW_2020}
Niccodemi, G., R.~Alessie, V.~Angelini, J.~Mierau, and T.~Wansbeek (2020).
\newblock Refining clustered standard errors with few clusters.
\newblock Working Paper 2020002-EEF, University of Groningen.

\bibitem[\protect\citeauthoryear{Pustejovsky and Tipton}{Pustejovsky and
  Tipton}{2018}]{PT_2018}
Pustejovsky, J.~E. and E.~Tipton (2018).
\newblock Small sample methods for cluster-robust variance estimation and
  hypothesis testing in fixed effects models.
\newblock {\em Journal of Business \& Economic Statistics\/}~{\em 36},
  672--683.

\bibitem[\protect\citeauthoryear{Roodman, Mac\-Kinnon, Nielsen, and
  Webb}{Roodman et~al.}{2019}]{RMNW}
Roodman, D., J.~G. Mac\-Kinnon, M.~{\O}. Nielsen, and M.~D. Webb (2019).
\newblock Fast and wild: Bootstrap inference in {Stata} using boottest.
\newblock {\em Stata Journal\/}~{\em 19}, 4--60.

\bibitem[\protect\citeauthoryear{Young}{Young}{2022}]{AY-IV}
Young, A. (2022).
\newblock Consistency without inference: Instrumental variables in practical
  application.
\newblock {\em European Economic Review\/}~{\em 147}, 1--21.

\end{thebibliography}
\addcontentsline{toc}{section}{\refname}

\end{document}